\documentclass[twocolumn]{29onr_art}
\usepackage[round]{natbib}
\usepackage{ulem}
\usepackage{graphicx}
\usepackage{fixltx2e}
\usepackage{amsmath,float}
\usepackage{graphicx}
\usepackage{amsmath,amssymb}
\usepackage[english]{babel}
\usepackage{epsfig}
\usepackage{color}
\usepackage{epstopdf}
\usepackage{dblfloatfix}
\usepackage[colorlinks=true,
            urlcolor=blue,
            linkcolor=docgreen,
            citecolor=docgreen,
            bookmarks=true,
            pdftitle={Near-Surface Boundary Layer Turbulence Along a Horizontally-Moving, Surface-Piercing Vertical Wall},
            pdfauthor={Nathan Washuta},
            pdfsubject={Paper submitted for the 31st Naval Hydrodynamics Symposium},
            pdfkeywords={Boundary layer, free surface},
	    bookmarksopen=false,
            pdfpagemode=None]{hyperref}

\definecolor{docgreen}{rgb}{0,0.5,0}

\pdfoutput = 1

\begin{document}

\title{Near-Surface Boundary Layer Turbulence Along a Horizontally-Moving, Surface-Piercing Vertical Wall}

\author{
N. Washuta, N. Masnadi, and J. H. Duncan}

\affiliation{\small University of Maryland, College Park, Maryland, USA}

\maketitle

\begin{abstract}

The complex interaction between turbulence and the free surface in boundary layer shear flow created by a vertical surface-piercing wall is considered. A laboratory-scale device was built that utilizes a surface-piercing stainless steel belt that travels in a loop around two vertical rollers, with one length of the belt between the rollers acting as a horizontally-moving flat wall. The belt is accelerated suddenly from rest until reaching constant speed in order to create a temporally-evolving boundary layer analogous to the spatially-evolving boundary layer that would exist along a surface-piercing towed flat plate. Surface profiles are measured with a cinematic laser-induced fluorescence system and sub-surface velocity fields are recorded using a high-speed planar particle image velocimetry system. It is found that the belt initially travels through the water without creating any significant waves, before the free surface bursts with activity close to the belt surface. These free surface ripples travel away from the belt before appearing to become freely-propagating waves. From sub-surface velocity measurements, it is found that close to the surface, transition to turbulence happens sooner than far from the surface, leading to an overall thicker boundary layer in the vicinity of the free surface. A secondary peak in streamwise velocity fluctuations accompanies this transition to turbulence and this peak reaches a maximum value a short time later before smoothing outward. Using momentum thickness as a length scale and the streamwise velocity fluctuations at the location of this outer peak as a velocity scale, free surface bursting and air entrainment onset are found to depend in some way on Weber number and agreement is found with scaling arguments for air entrainment presented by \citet{broc:2001}.

\end{abstract}

\section{Introduction}

\begin{figure}
\begin{center}
\includegraphics[trim=0 0.1in 0 0.1in, clip=true,width=3in]{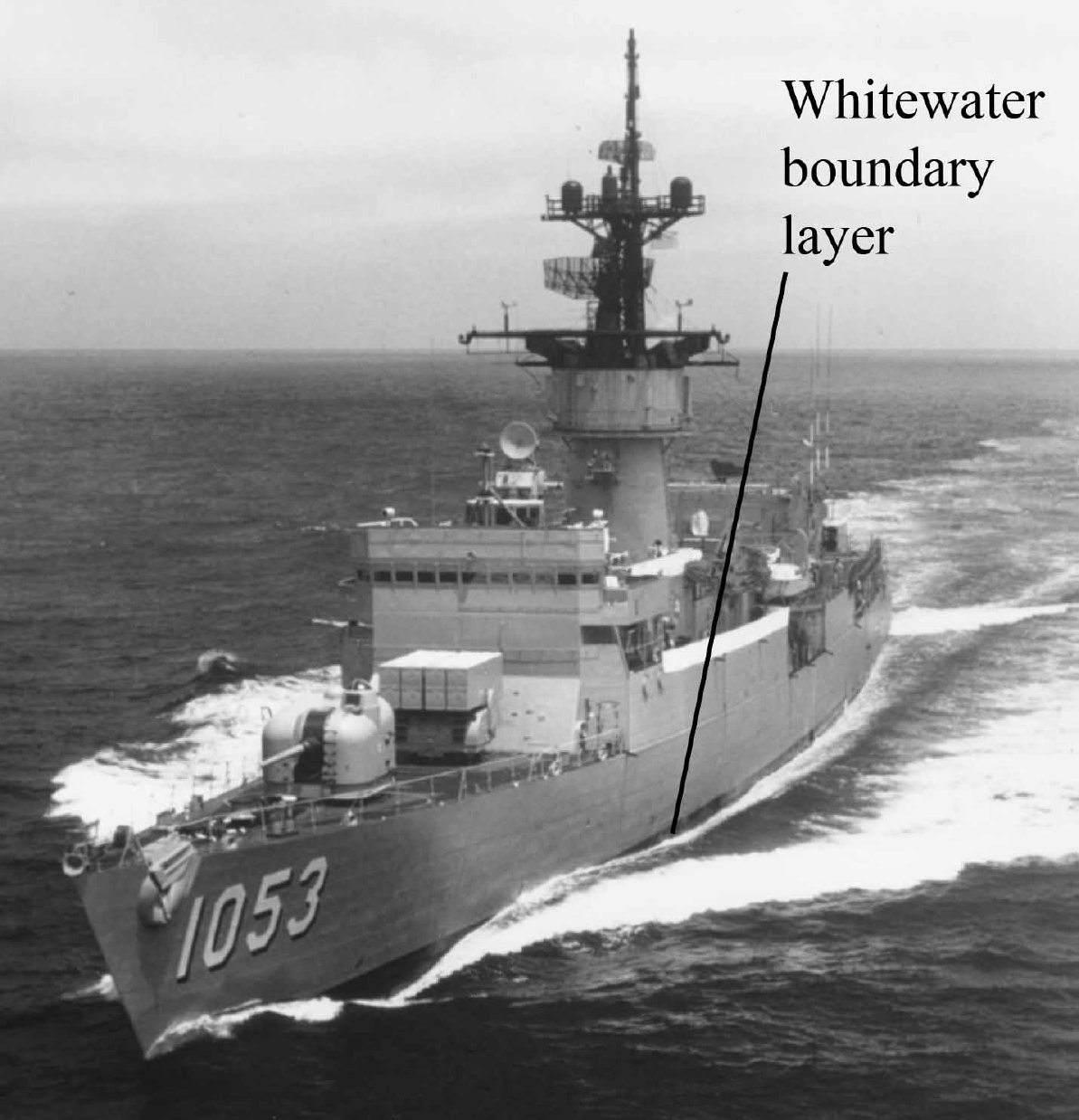}
\end{center}
\vspace*{-0.17in} \caption{Photograph of naval combatant ship showing zone of white water next to the hull.} \label{fig:ship}
\end{figure}

Turbulent boundary layers near the free surface along ship hulls and surface-piercing flat plates have been explored by a number of authors, see for example \citet{Longo1998}, \citet{Sreedhar1998}, \citet{Stern1989} and \citet{Stern1993}. However, even though it has long been observed that there is a layer of white water next to the hulls of naval combatant ships moving at high speed, see for example the photograph in Figure~\ref{fig:ship}, the entrainment of air at the free surface in ship boundary layers has received relatively little attention. It is not known whether this white water is the result of active spray generation and air entrainment due to turbulence in the boundary layer along the ship hull or the result of spray and air bubbles that are generated upstream in the breaking bow wave and then swept downstream with the flow. In the free surface boundary layer, the air entrainment process is controlled by the ratios of the turbulent kinetic energy to the gravitational potential energy and the turbulent kinetic energy to the surface tension energy. The ratio of the turbulent kinetic energy to the gravitational potential energy is given by the square of the turbulent Froude number ($Fr^2 = q^2 / (g L)$) and the ratio of turbulent kinetic energy to surface tension energy is given by the Weber number ($We = \rho q^2 L/ \sigma$), where $g$ is the acceleration of gravity, $\rho$ is the density of water, $\sigma$  is the surface tension of water, $q$ is the characteristic magnitude of the turbulent velocity fluctuations and $L$ is the length scale of this turbulence. 

\begin{figure}
\begin{center}
\includegraphics[width=3in]{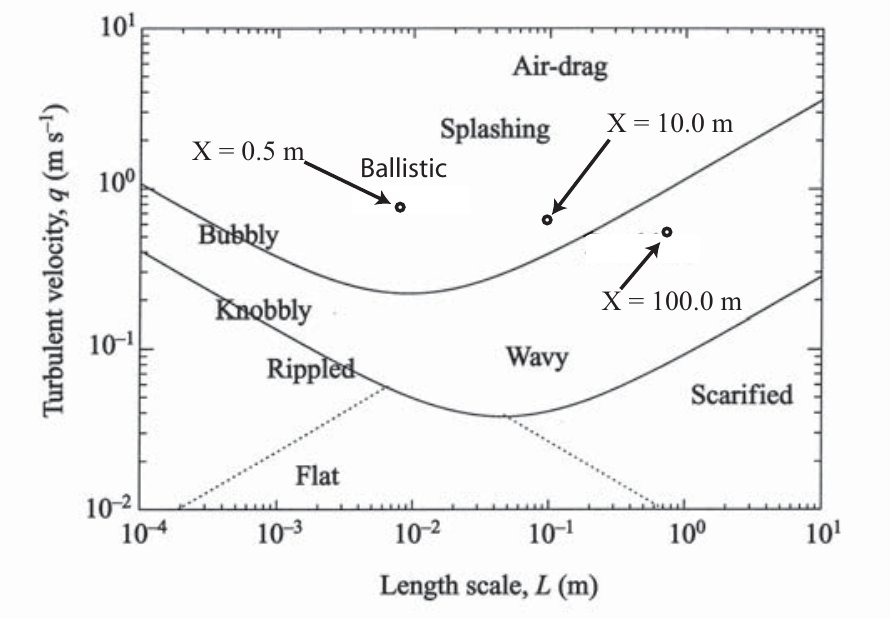}
\end{center}
\vspace*{-0.2in} \caption{Regions of various types of surface motions for free surface turbulence with velocity fluctuation magnitude $q$ (vertical axis) and length scale $L$ (horizontal axis), from \citet{broc:2001}. Air entrainment and spray production occur in the upper region, above the uppermost curved line. The three data points are values obtained for the turbulent boundary layer on a flat plate with $q$ taken as the rms of the spanwise (which is vertical for the boundary layer along a ship hull) velocity fluctuation ($w'$) and L taken as the boundary layer thickness ($\delta$).} \label{fig:brocchiniperegrine}
\vspace{-0.04in}
\end{figure}

Several authors have applied theory and numerical methods to explore the interaction of turbulence and a free surface, see for example \citet{Shen2001}, \citet{Guo2009}, \citet{kim:2013} and \citet{broc:2001}. \citet{broc:2001} have used scaling arguments to predict the critical Froude and Weber numbers above which air entrainment and spray generation will occur due to strong free-surface turbulence. Figure~\ref{fig:brocchiniperegrine}, which is from their paper, shows the boundaries of various types of surface undulations on a plot of $q$ versus $L$. The upper region of the plot is the region of air entrainment and droplet generation. We have used classical boundary layer correlations to make estimates of $q$ (taken as the root-mean-square vertical component of the turbulent velocity fluctuations) and $L$ (taken as the boundary layer thickness) at three streamwise positions in a ship boundary layer and plotted these points on the $q$-$L$ map in Figure~\ref{fig:brocchiniperegrine}. While these are simply estimates and perhaps more appropriate turbulent length scales can be considered, these estimates serve as an adequate first guess for the behavior of this flow. As can be seen from the figure, the points are clearly in the air entrainment region of the plot, especially the points near the bow. Thus, air entrainment due to strong turbulent fluid motions in the hull boundary layer at the free surface is a likely cause of the layer of white water.

The difficulty with laboratory experiments on bubble entrainment and spray stems from the fact that the experiments are performed in the same gravitational field as found in ship flows and that the only practical liquid available is water, as is also found in the ocean. Thus, with $g$, $\rho$ , and $\sigma$ the same in the field and in the laboratory, one must attempt to achieve full-scale flow speeds in order to obtain Froude, Weber, and Reynolds similarity with field conditions. Also, even if full scale-values of $q$ and $L$ were obtained by towing a surface piercing flat plat with the length of the ship at high speed in a ship model basin, the free surface flow would include a bow wave which would obfuscate the source of the bubbles and spray. Another problem is that in order to obtain realistic entrainment/spray conditions and bubble/droplet size distributions, these experiments should be performed in salt water which is not typically used in ship model basins.

In view of the above difficulties in simulating air entrainment due to the turbulent boundary layer, we have built a novel device that produces an approximation of a full scale ship boundary layer in the laboratory. This device, called the Ship Boundary Layer (SBL) simulator generates a temporally evolving boundary layer on a vertical, surface-piercing flat wall. This vertical wall consists of a stainless steel belt loop that is 1.0~m wide and about 15 m long. The belt is mounted on two vertically oriented rollers as shown in Figure~\ref{fig:TankSchem}. The rollers are driven by hydraulic motors and the entire device is placed in a large open-surface water tank as shown in the figure. Before each experimental run, the belt and the water in the tank are stationary. The water level is set below the top edge of the belt and the flow outside the belt loop on one of the long lengths between the rollers is studied. The belt is accelerated from rest using a hydraulic control system, which is able to reproduce the same belt motion within 10~cm of travel over the length of a run.

In the experiments discussed herein, the belt is accelerated suddenly from rest until it reaches a pre-defined speed which is held steady for a short time. The flow on the surface of the belt in this case is a simulation of the flow seen by a stationary observer in the ocean as a ship, that makes no waves, passes by at constant speed. The temporally-evolving boundary layer created along the entire length of the belt can be considered equivalent to the spatially-developing boundary layer along a flat ship hull, with the distance along the ship hull corresponding to any time $t$ after the belt begins to move is essentially the distance traveled by the belt, $x$. 

The remainder of this paper is divided into three sections.  The experimental setup is described first. This is followed by the presentation and discussion of the results. Conclusions of this study are given in the end.

\section{Experimental Details}

\begin{figure*}[!htb]
\begin{center}
\begin{tabular}{cc}
\includegraphics[trim=0 0.6in 0 0.00in,clip=true,height = 1.6in]{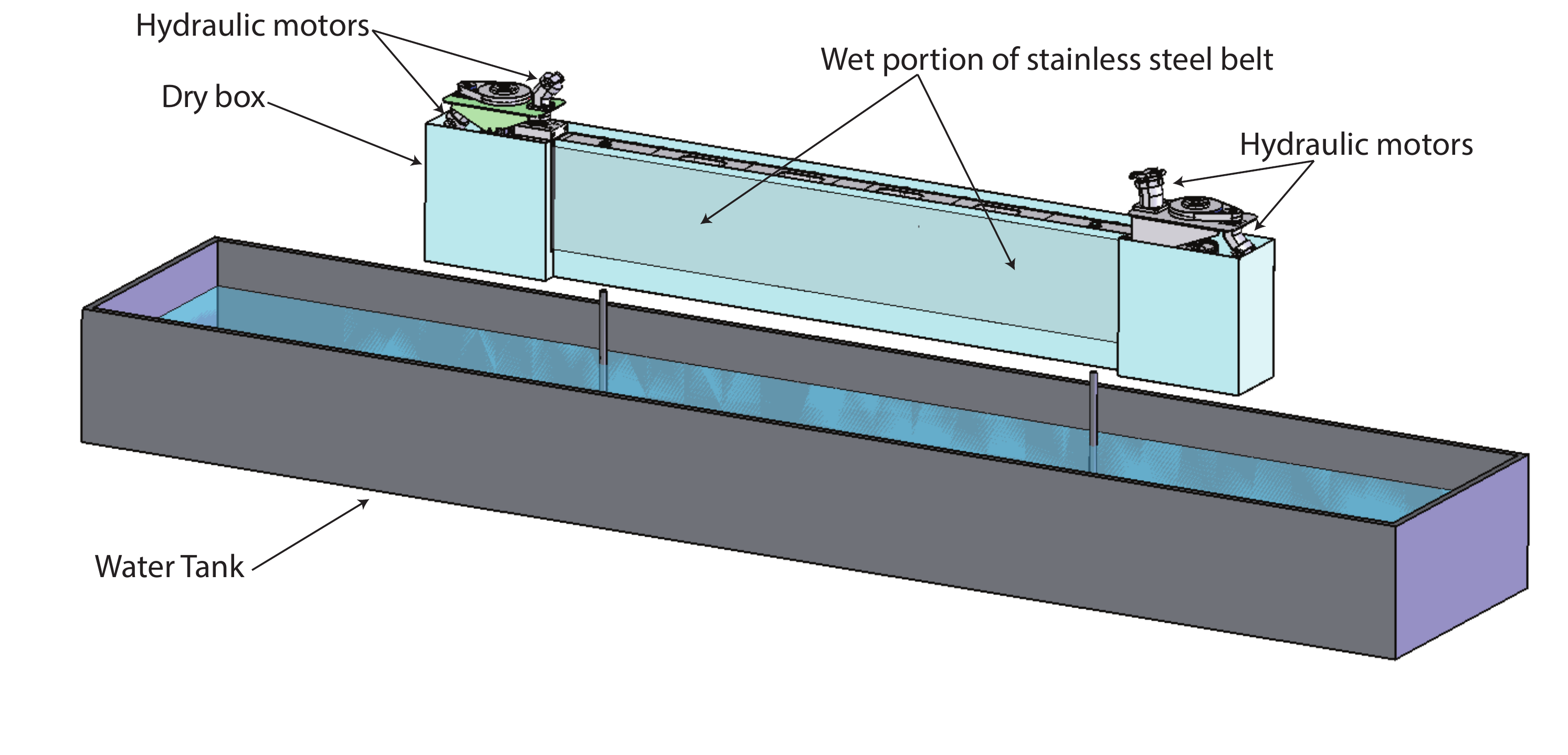}
\includegraphics[trim=0in 1.2in 0 0.00in,clip=true,height = 1.6in]{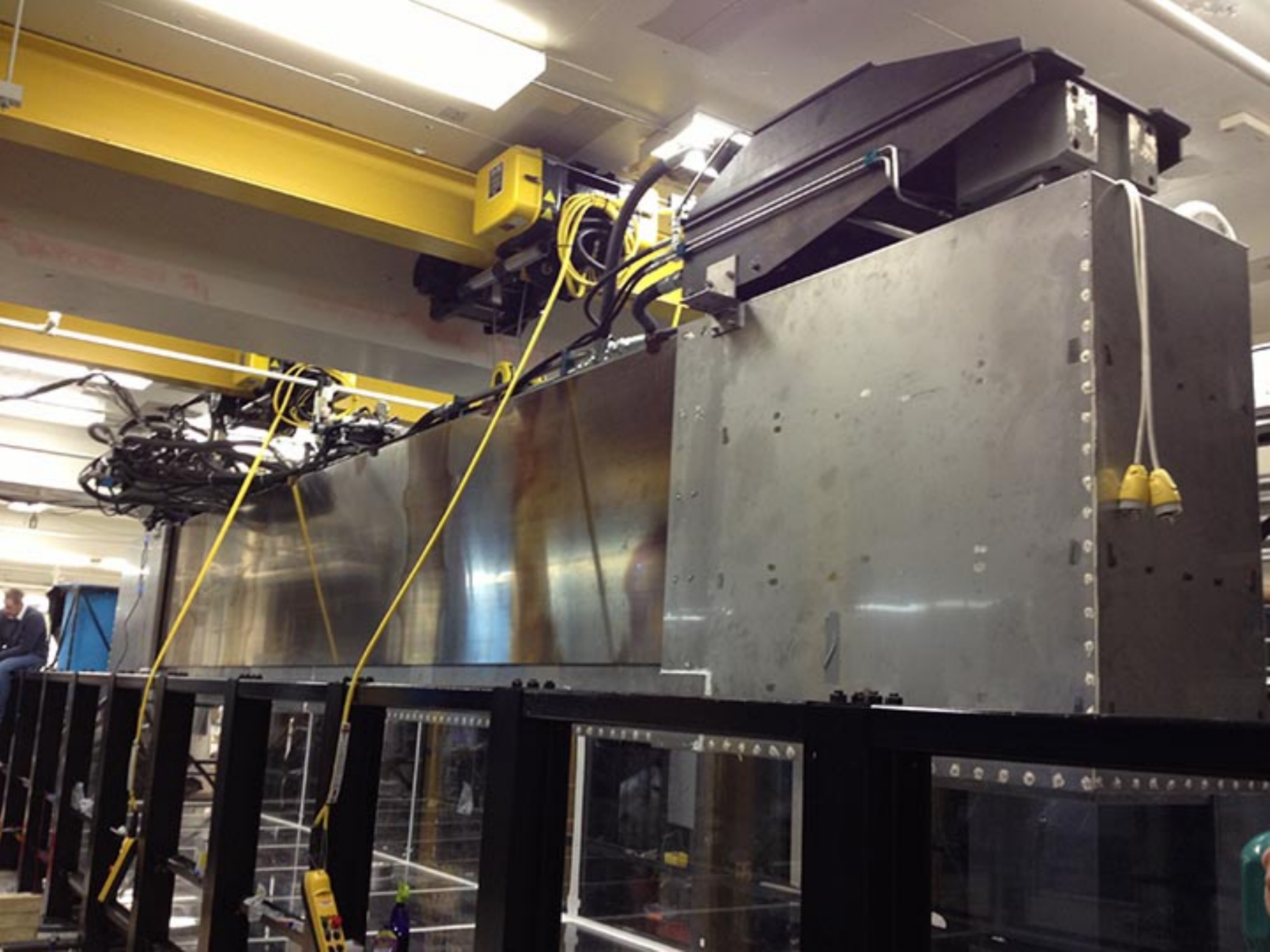} 
\end{tabular}
\end{center}
\vspace*{-0.1in} \caption{Perspective view of the Ship Boundary Layer Simulator (SBLS) and the water tank and a photo of the belt device being lowered into the tank.} \label{fig:TankSchem}
\end{figure*}

The experiments are performed in an open-surface water tank that is 13.34 m long, 2.37 m wide and 1.32 m deep, see Figure~\ref{fig:TankSchem}.  The tank structure is made up of steel beams which support 31.8-mm-thick clear Acrylic panels that make up the floor and walls of the tank and provide optical access from beneath the water surface. The top of the tank is open, offering an unobstructed view of the water surface. Two floor-mounted circular steel pads pierce the bottom of the tank and are used to support the Ship Boundary Layer (SBL) simulator. The tank includes a water filtration system consisting of a 2.3-m-long skimmer at one end of the tank, a diatomaceous-earth filter, a centrifugal pump and associated pipes and valves.  The output line of the skimmer can be directed either to the drain or to the filter where the water is returned to the opposite end of the tank.

\begin{figure*}[!htb]
\begin{center}
\includegraphics[trim=0 0.0in 0in 0in,clip=true,width = 5in]{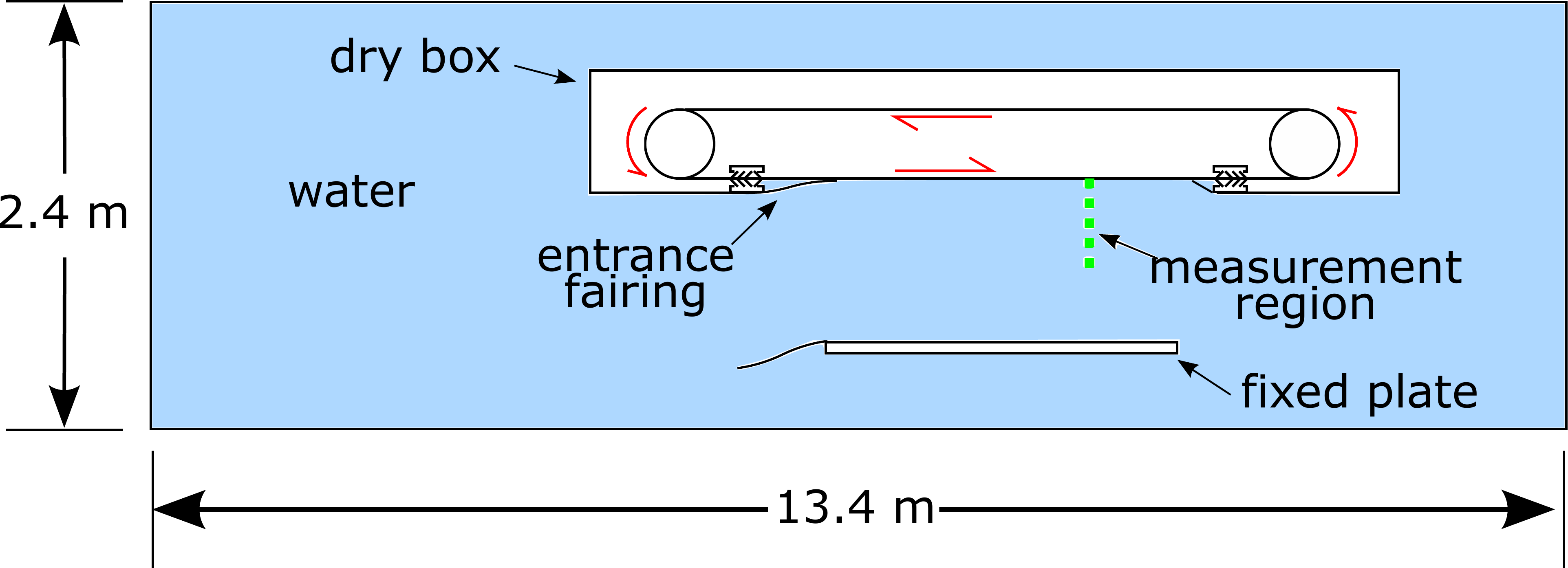}
\end{center}
\vspace*{-0.2in} \caption{Plan view of the SBL and water tank.} \label{fig:top_view}
\end{figure*} 

The main functional component of the SBL is a one-meter-wide 0.8-mm-thick endless stainless steel belt which is driven by two 0.46-meter-diameter, 1.1-meter-long rollers whose  rotation axes are vertically oriented and separated by a horizontal distance of approximately 7.5 meters. The rollers are each driven by two bent-axis hydraulic motors via toothed belt and pulley systems.  Each roller along with the motors and drive systems form single drive units that are attached to a welded steel frame that maintains the separation between and relative parallel orientation of the rollers. The roller drive unit on the left in Figure~\ref{fig:TankSchem} is attached to the steel frame via two hydraulic pistons, positioned at the top and bottom of the frame.  The vertical position of the belt  on the roller is controlled actively during each experimental  run by measuring the belt position with a light sensor and tilting the left roller with differential motion of the hydraulic pistons.  Tilting the left roller clockwise (counter clockwise) by small amounts causes the belt to move up (down).  During a typical run, the roller position varies by no more than $\pm 2.5$~mm.

The assembled SBL device is placed in a stainless steel sheet metal box (called the dry box) as shown in Figures~\ref{fig:TankSchem} and \ref{fig:top_view}.  The box keeps the assembly essentially dry, while one of the two straight sections of the belt exits the box through a set of seals near the roller on the left and travels through the water to the second set of seals near the opposite roller where the belt re-enters the box. This box was deemed necessary to keep water from corroding the SBL mechanism and to keep water from being dragged in between the belt and the roller where it might cause the belt to hydroplane off the roller. The lone straight section exposed to water is approximately 6 meters long and pierces the free surface with approximately 0.33 meters of freeboard for the water level used in the present experiments. At the location where the belt leaves the dry box and enters the water, a sheet metal fairing is installed to reduce the flow separation caused by the backwards-facing step associated with the shape of the dry box at this location.

\begin{figure*}[!htb]
\begin{center}
\begin{tabular}{cc}
\includegraphics[trim=0.24in 0.0in 0 0.00in,clip=true,scale=0.6]{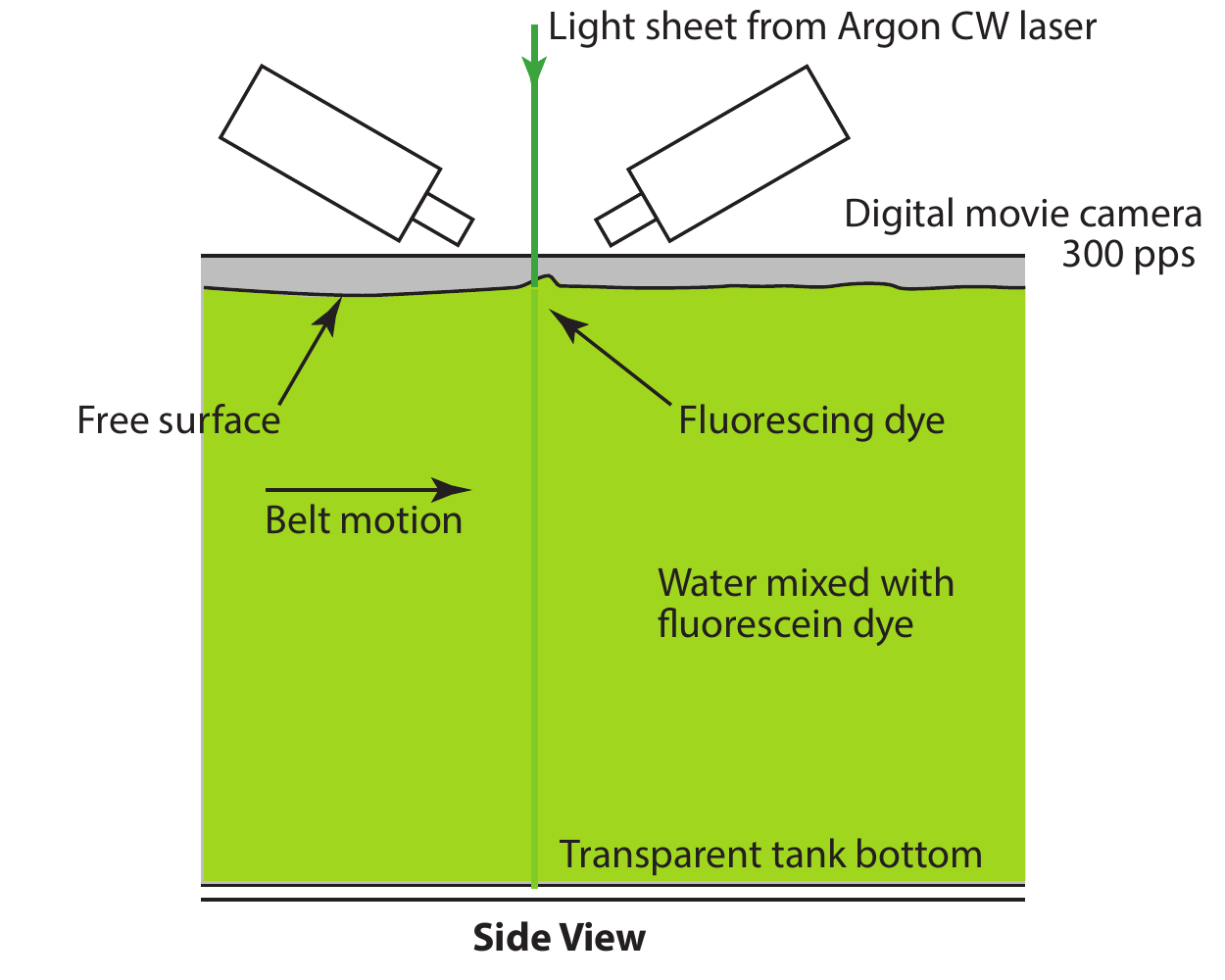}
\includegraphics[trim=0in 0.0in 0 0.00in,clip=true,scale=0.6]{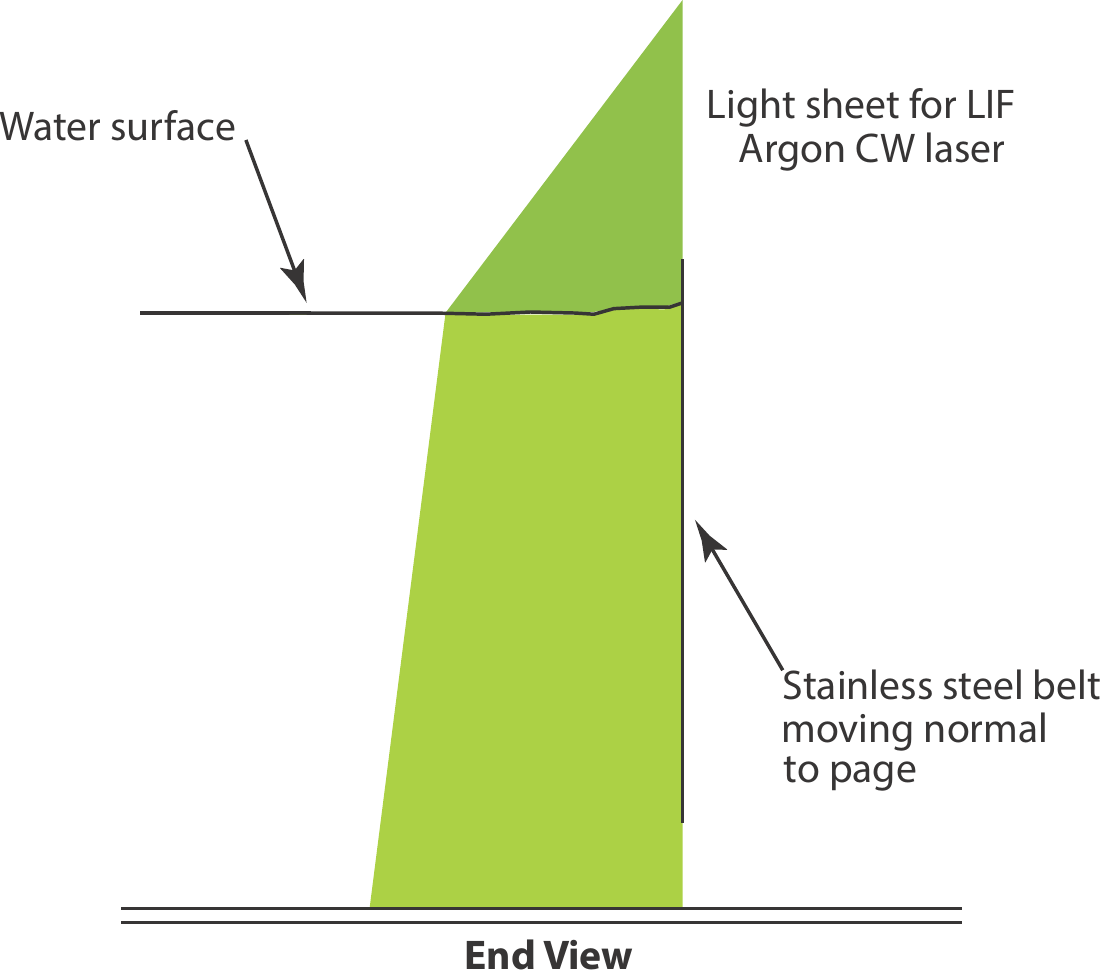}
\end{tabular}
\end{center}
\vspace*{-0.1in} \caption{Schematic drawing showing the set up for the cinematic LIF measurements of the free surface shape.} \label{fig:LIFSchem}
\end{figure*} 

\begin{figure*}[!ht]
\begin{center}
\begin{tabular}{cc}
\includegraphics[trim=0.0in 0.0in 0 0.00in,clip=true,width = 2.5in]{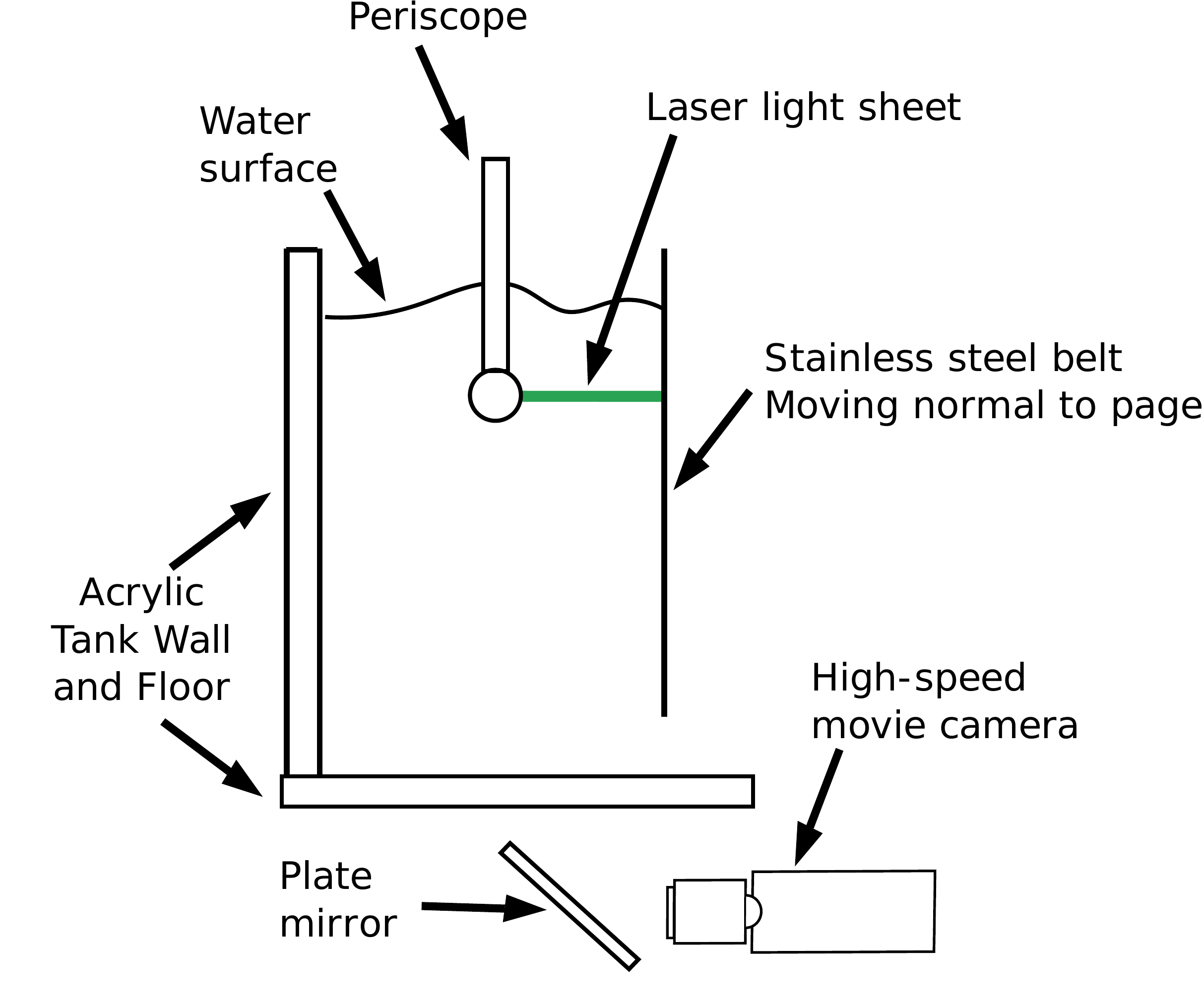} &
\includegraphics[trim=0in 0.0in 0 0.00in,clip=true,width = 3.5in]{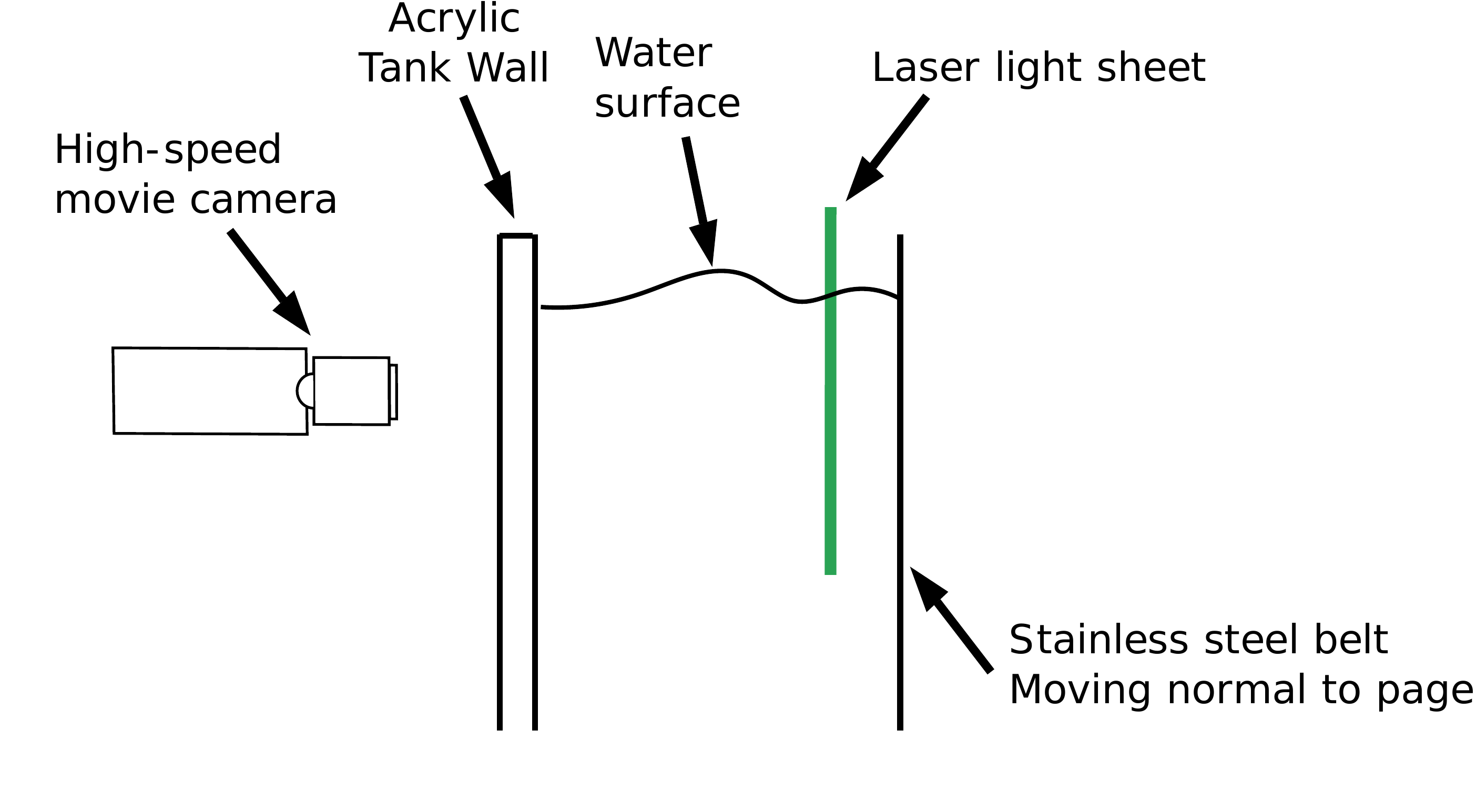}\\
(\textit{a}) & (\textit{b})
\end{tabular}
\end{center}
\vspace*{-0.1in} \caption{Schematic drawing showing the two configurations of the planar PIV setup with (\textit{a}) a horizontal light sheet and (\textit{b}) a vertical light sheet.} \label{fig:PIV}
\end{figure*} 

When performing experiments, the belt is launched from rest and accelerates until reaching constant speed. Throughout these transient experiments, the belt travel is analogous to the passage of a flat-sided ship that makes no bow waves; the length along the hull is equivalent to the total distance traveled by the belt. Belt speeds ranging from 3 to 5~m/s were used and measurements were continued until a  belt length of 30~m had passed by the measurement site. Therefore, at 3~m/s, experiments run for 10 seconds, while at 5~m/s, experiments run for 6 seconds. While the time to accelerate varies depending on belt speed, independent measurements of belt travel show that during launch the belt travels 0.85, 1.45, and 2.29~m at belt speeds of 3, 4, and 5~m/s, respectively.

To study the water surface deformation, a cinematic Laser Induced Fluorescence (LIF) technique was  utilized, see Figure~\ref{fig:LIFSchem}. In this technique, a continuous-wave Argon Ion laser beam is converted to a thin sheet using a system of spherical and cylindrical lenses. This sheet is projected vertically down onto the water surface in an orientation with the plane of the light sheet normal to the plane of the belt. This laser emits light primarily at wavelengths of 488~nm and 512~nm. The water in the tank is mixed with fluorescein dye at a concentration of about 5~ppm and dye within the light sheet fluoresces. Two cameras view the intersection of the light sheet and the water surface from both upstream and downstream with  viewing angles of approximately 20 degrees from  horizontal. The cameras (Phantom V641 by Vision Research, Inc.) capture 4-Mpixel 12-bit black-and-white images at frame rates of 1000~Hz.  A long-wavelength-pass optical filter is placed in front of each camera lens. These filters block out the laser light and transmit the light from the fluorescing dye, thus preventing specular reflections of the laser light from the water surface from entering the camera lenses. The use of two cameras allows for more accurate surface measurement in the event that the intersection of the light sheet and the water surface is blocked  by large deformations of the free surface between the plane of the light sheet and either camera. The images seen by these cameras shows a sharp line at the intersection of the light sheet with the free surface. Using image processing, instantaneous surface profiles can be extracted from these images.

In addition to the above-described surface profile measurements, sub-surface velocity fields were measured in planes below the free surface using cinematic planar Particle Image Velocimetry (PIV), as can be seen in Figure~\ref{fig:PIV}. In this technique, the flow is seeded homogeneously with fluorescent tracer particles which have sufficient size and density to faithfully follow the flow. The beam from a high-repetition-rate laser is formed into a thin light sheet and projected from below the free surface, illuminating tracer particles in a plane. A high-speed camera captures image pairs of these illuminated tracer particles with a short time separation between frames. By performing cross-correlation between interrogation windows of successive frames, two-component velocity vector fields can be determined at each instant in time throughout a run. In these experiments, two configurations of this PIV setup are utilized. In the first, the light sheet is oriented horizontally and projected from a periscope directly towards the belt surface. A camera under the tank views a plate mirror, which directs its view upward through the floor of the tank, recording streamwise and wall-normal velocity components. Light sheet depths, $D$, of 2.5~cm and 14~cm below the undisturbed free surface are utilized. In the second configuration, the periscope is moved closer to the belt surface and reoriented so that the light sheet is projected upward in a plane parallel to the belt surface. Light sheet locations of 0.75 and 1.5~cm from the belt surface are utilized.

\section{Results and Discussion}

\subsection{Surface Profiles}

\begin{figure}[!ht]
\begin{center}
\begin{tabular}{c}
\includegraphics[trim=0 0.0in 0 0.00in,clip=true,width=3in]{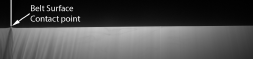}\\
(\textit{a})\\
\includegraphics[trim=0 0.0in 0 0.00in,clip=true,width=3in]{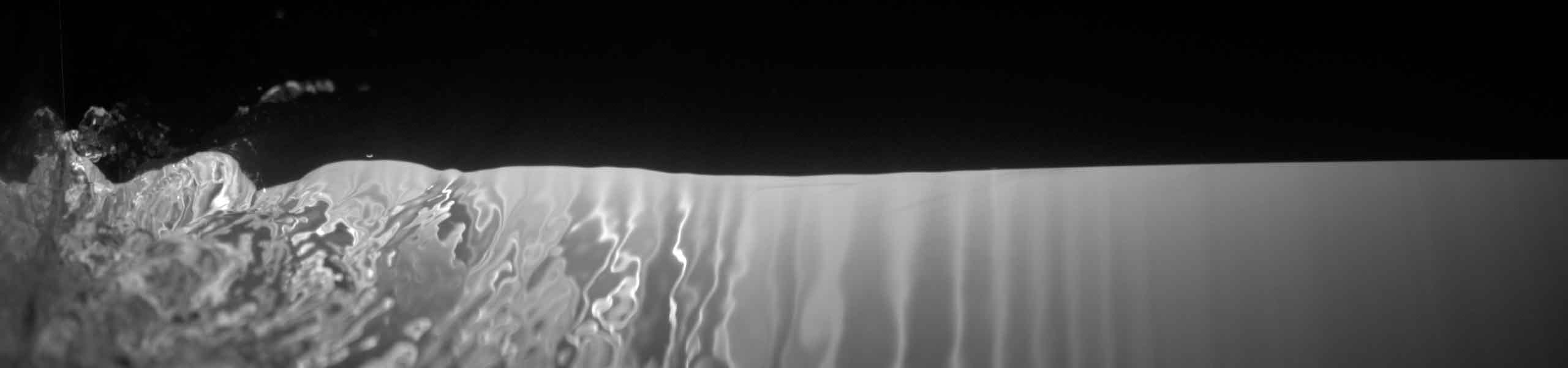}\\
(\textit{b})\\
\includegraphics[trim=0 0.0in 0 0.00in,clip=true,width=3in]{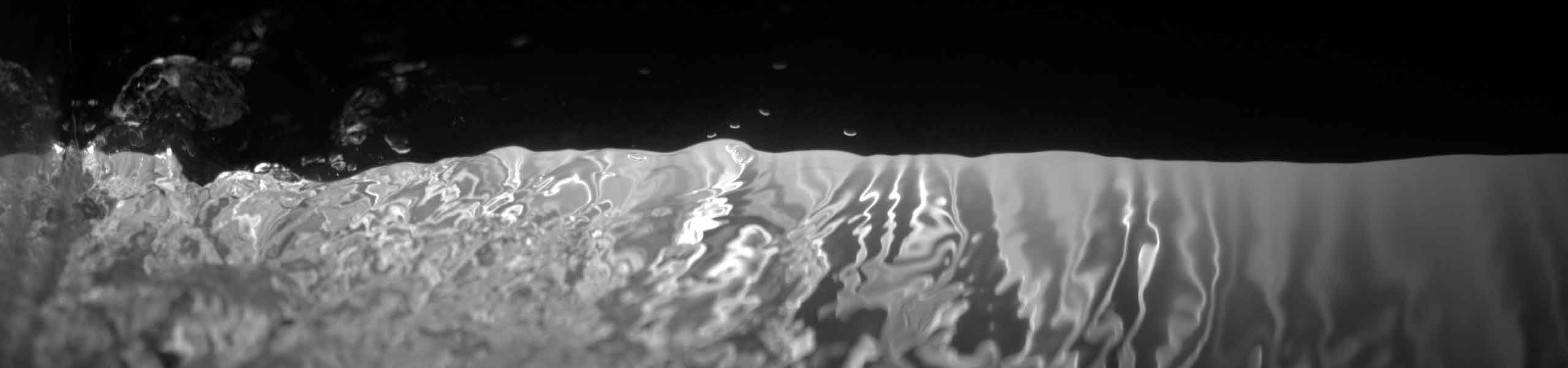}\\
(\textit{c})\\
\includegraphics[trim=0 0.0in 0 0.00in,clip=true,width=3in]{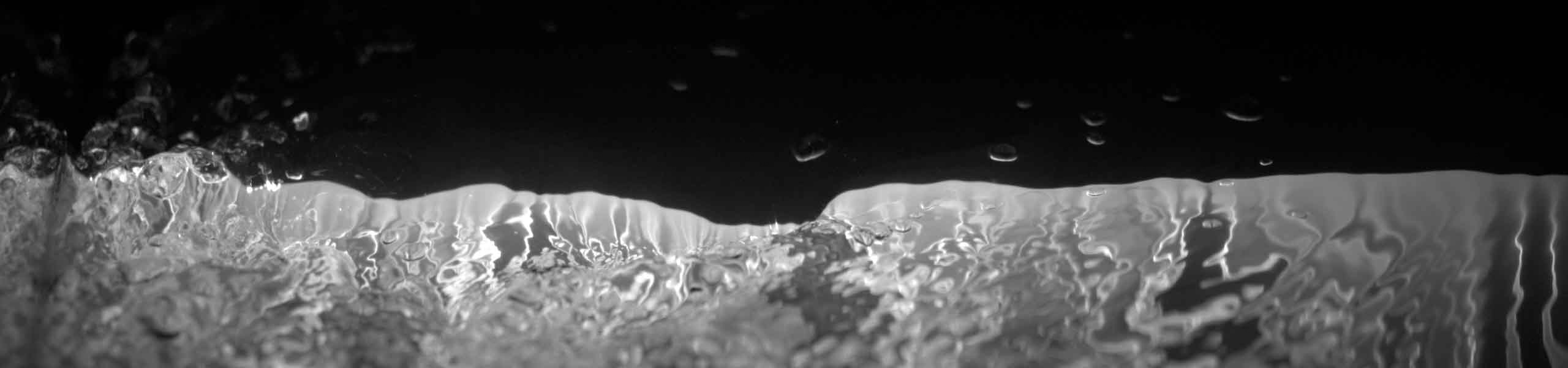}\\
(\textit{d})\\
\includegraphics[trim=0 0.0in 0 0.00in,clip=true,width=3in]{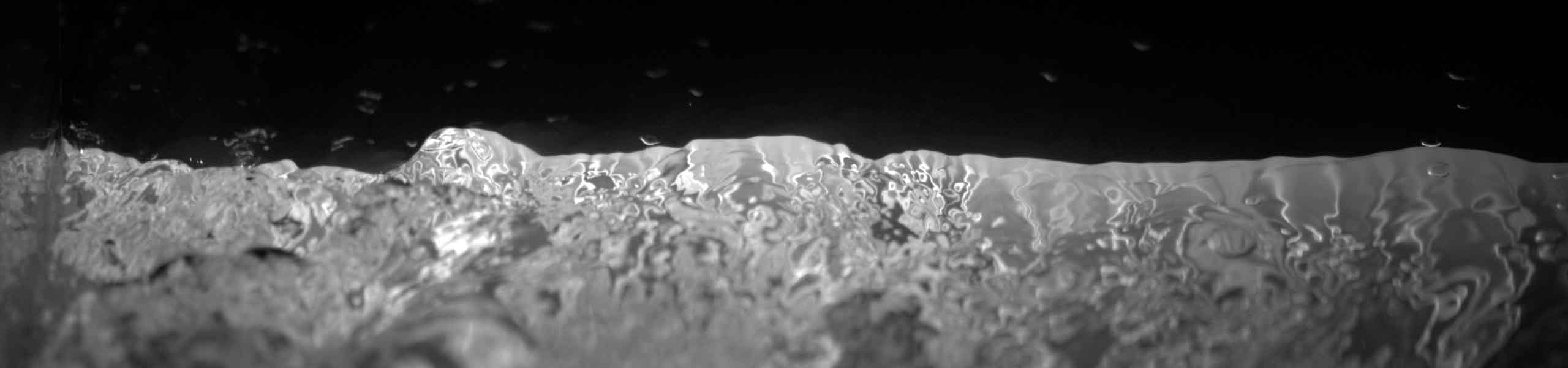}\\
(\textit{e})\\
\end{tabular}
\end{center}
\vspace*{-0.2in} 
\caption{A sequence of five images from a high speed movie of the free surface during a belt launch to 5~m/s. These images are taken at equivalent belt lengths of (\textit{a}) 0~m (\textit{b}) 5~m (\textit{c}) 10~m (\textit{d}) 15~m and (\textit{e}) 20~m from the bow of the ship. The high reflectivity of the stainless steel belt makes it appear as a symmetry plane on the left side of the images. The horizontal field of view for these images is approximately 31~cm.}\label{fig:overall}
\end{figure}

In this set of experiments, surface profile measurements were performed at belt speeds, $U$, of 3, 4, and 5~m/s. Through initial trials, it was determined that a frame rate of 1000~fps was necessary to provide a sufficient temporal resolution so that surface features could be identified and tracked smoothly in successive frames. LIF images of the water free surface next to the belt for an experimental run with $U$ = 5~m/s are shown in Figure~\ref{fig:overall}. The five images in the figure are spaced out equally by distance of belt travel, with the first image (\textit{a}) taken at 0.0~s, the time when the belt first starts to move. The instantaneous belt speed from the beginning of belt motion through the acceleration portion until the belt reaches constant speed has been measured separately and is used to correlate the time of each frame to the belt travel distance. Here and in the following, rather than refer to images and data by the time after the belt has started moving, we refer to them by the distance, $x$, from the leading edge of an equivalent flat plate, which is also the distance that the belt has traveled
\[
x = \int_0^t U(t)dt.
\]
 Thus, the images in Figure~\ref{fig:overall} depict a portion of a run, with images (\textit{a}), (\textit{b}), (\textit{c}), (\textit{d}) and (\textit{e}) captured at 0~s, 1.35~s, 2.35~s, 3.35~s, and 4.35~s, respectively, corresponding to $x=$ 0.0, 5.0, 10.0, 15.0 and 20.0~m. 

As discussed in the previous section, the plane of the vertical light sheet is oriented normal to the belt surface and the cameras look parallel to the belt surface and down at the water surface at a small angle from horizontal. The images in Figure~\ref{fig:overall} are from the downstream camera, and these images have been flipped horizontally for convenience in order to match the coordinate system of later plots, so that the belt is near the left side of each image and is moving out of the page. The position of the belt is marked on the left side of image (\textit{a}) and the intensity pattern to the left of this location is a reflection of the light pattern on the right due to the high reflectivity of the smooth surface of the belt. This line of symmetry gives a good indication of the position of the belt in each image. The sharp boundary between the upper dark and lower bright region of each image is the intersection of the light sheet and the water surface. The upper regions of the later images contain light scattered from roughness features on the water surface behind the light sheet. These roughness features include bubbles that appear to be floating on the water surface and moving primarily in the direction of the belt motion. The bright area below the boundary is created by the glowing fluorescent in the underwater portion of the light sheet.  The complex light intensity pattern here is created by a combination of the refraction of the laser light sheet as it passes down through the water surface and the refraction of  the light from the glowing underwater dye as the light passes up through the water surface between the light sheet and the camera, on its way to the lens. The instantaneous shape of the free surface is extracted quantitatively from the intersection line of these images using gradient-based image processing techniques. It can be seen from these images that surface height fluctuations (ripples) are created close to the belt surface, at the left side of each image, and propagate away from the belt (to the right). As time passes, the surface height fluctuations grow dramatically and eventually surface breaking and air entrainment events begin to occur, resulting in bubble and droplet production.

\begin{figure}[!ht]
\begin{center}
\begin{tabular}{c}
(\textit{a})\\
\includegraphics[trim=0.0in 0.0in 0 0.00in,clip=true,width = 3in]{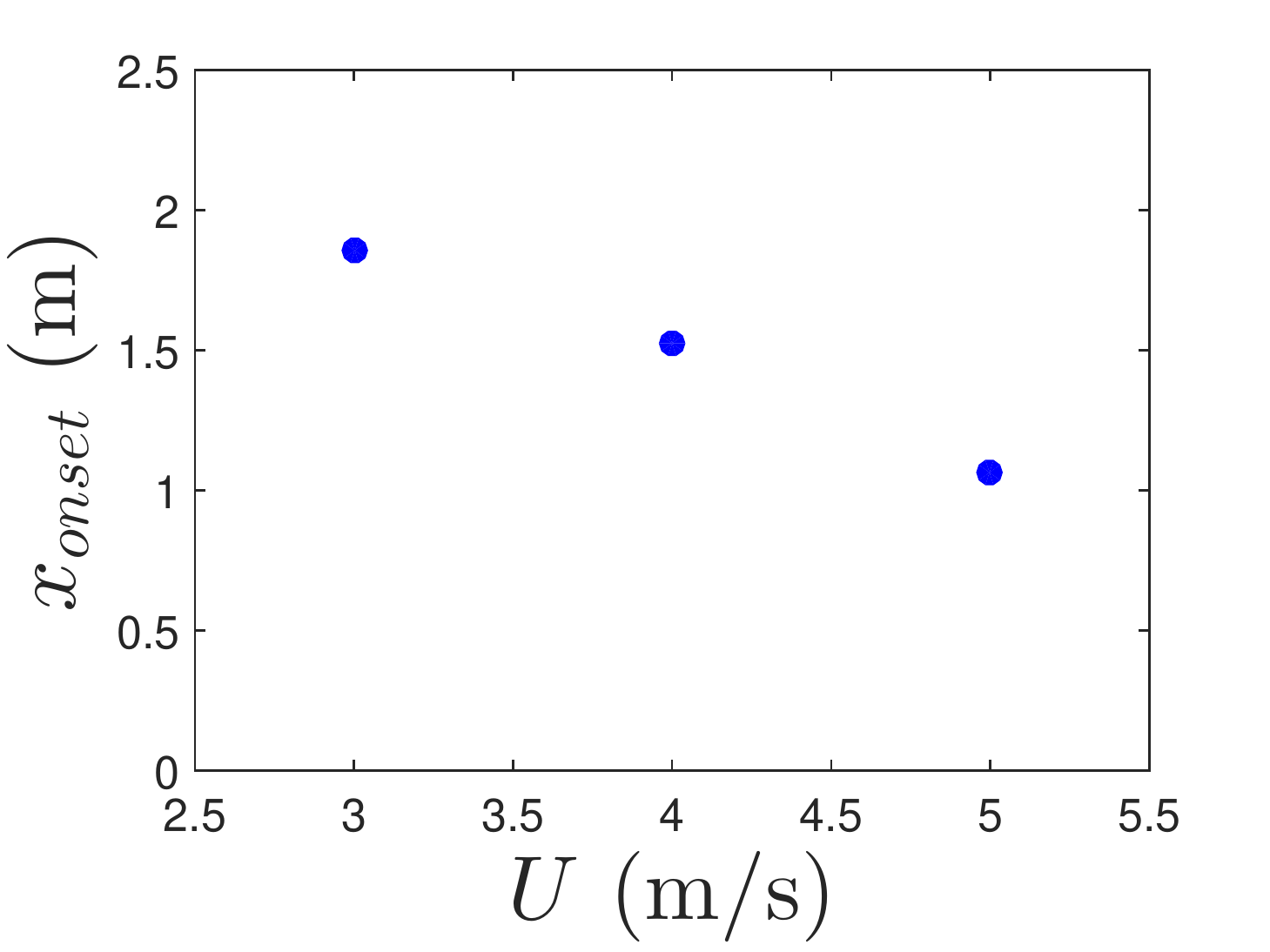}\\
(\textit{b})\\
\includegraphics[trim=0in 0.0in 0 0.00in,clip=true,width = 3in]{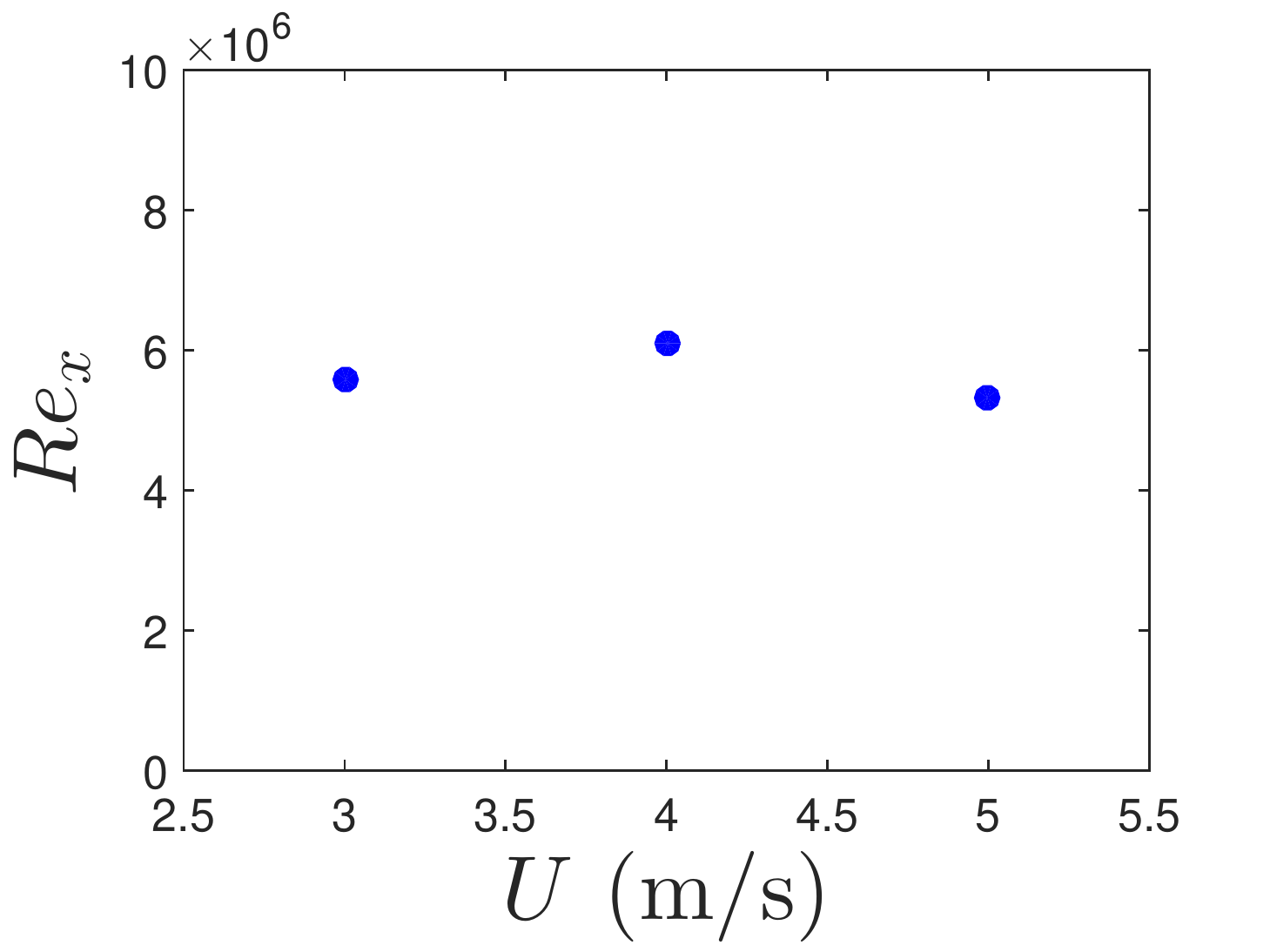}\\
\end{tabular}
\end{center}
\vspace*{-0.1in} \caption{Two plots showing (\textit{a}) the average $x$ location of the onset of free surface bursting versus belt speed and (\textit{b}) the associated Reynolds number based on $x$.} \label{fig:bursting}
\end{figure} 

\begin{figure*}[!htb]
\begin{center}
\includegraphics[trim=0 0in 0 0.5in, clip=true,scale=0.6]{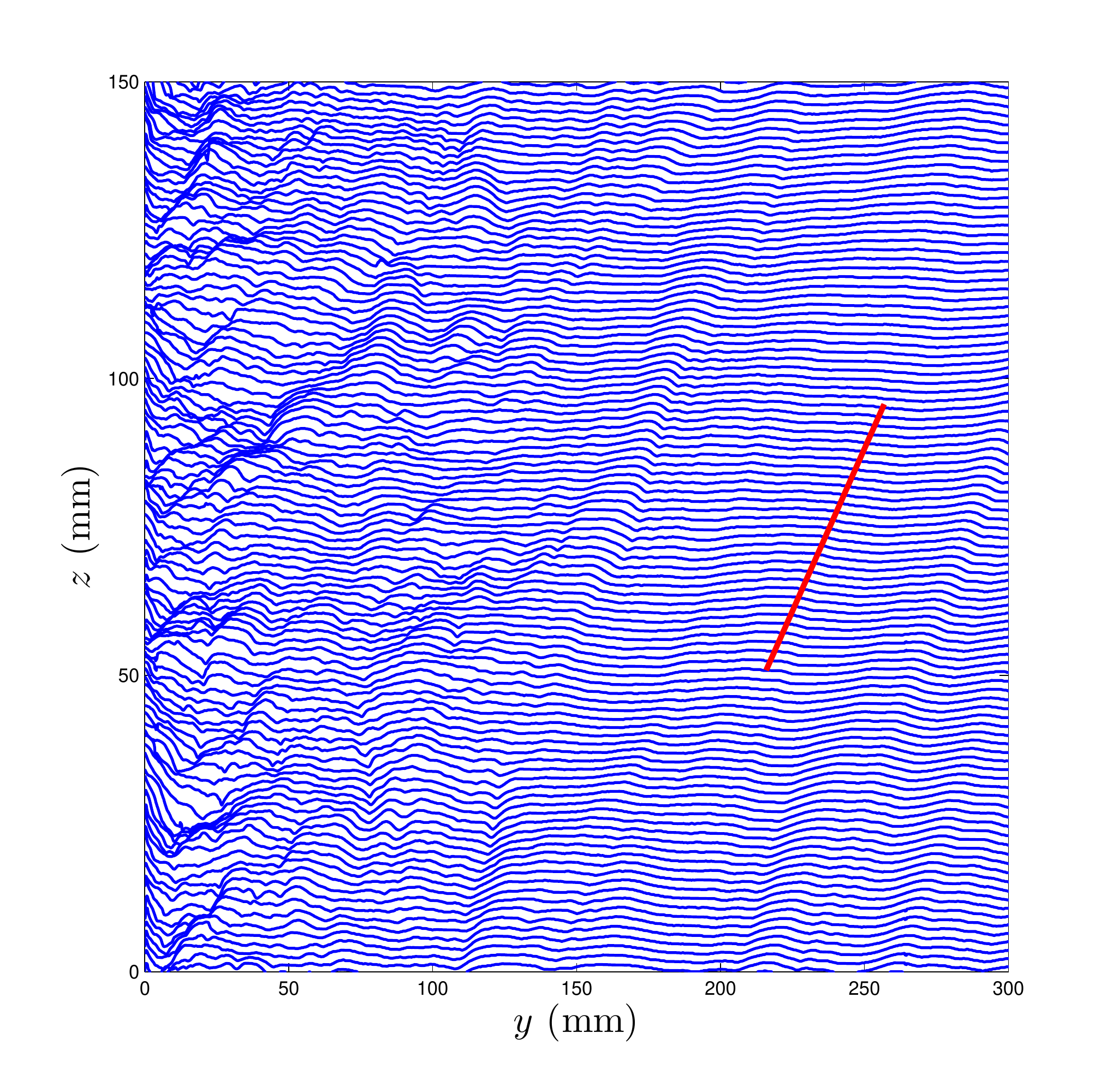}\\
\end{center}
\vspace*{-0.4in} \caption{Sequence of profiles of the water surface during belt launch to 3~m/s.  The time between profiles is 4~ms and each profile is shifted up 2~mm from the previous profile to reduce overlap and show propagation of surface features throughout time. The belt is positioned at the left edge of the image ($y =0$). The red line drawn over the image gives an estimation of the wall-normal propagation speed of surface features. In this case, the red line corresponds to a wall-normal velocity of approximately 34~cm/s.} \label{fig:profiles}
\end{figure*} 

In viewing the recorded LIF movies, it is clear that the free surface remains nearly quiescent during a period of belt travel at the beginning of each run; during this time period, the LIF images appear similar to what is seen in Figure~\ref{fig:overall}-(\textit{a}).  After a short time,  the surface suddenly bursts with activity near the belt surface, creating free surface ripples. After this point, see Figure~\ref{fig:overall}-(\textit{b}), the free surface fluctuations are continually generated close to the belt and this generation region grows in time. The sudden burst in surface activity may be an indication of transition to turbulence, but this bursting point could be a moderated turbulence signal that has been filtered by surface tension and gravity, which would act to restrain free surface motions. Therefore, the turbulence intensities may have to reach a critical threshold in order to mobilize the air-water interface. This will be further discussed in the following sections. The $x$ location of bursting onset is recorded in each LIF movie and the average over 20 runs at each belt speed is plotted versus $U$ in Figure~\ref{fig:bursting}-(\textit{a}).  As can be seen from the figure, there is a monotonic decrease of the onset location with increasing belt speed. The average Reynolds number based on the $x$ location of onset, is plotted versus $U$ in  Figure~\ref{fig:bursting}-(\textit{b}); the Reynolds number appears to be fairly consistent with an average value of $Re_x = 5.7 \times 10^6$. Other potential scaling parameters for this bursting onset will be discussed in more detail below.

In addition to qualitative observations of free surface motions, quantitative surface profiles can be extracted from each frame of the LIF movies. An example series of water surface profiles from a run with $U=3.0$~m/s over a range of $x$ from  21~m to 21.3~m is shown in Figure~\ref{fig:profiles}. The horizontal axis in the plot is horizontal distance, $y$, from the belt surface and the vertical axis is water surface height above the mean water level.  The profiles are equally spaced in $x$ by 1.2~cm and each successive profile is plotted 1.5~mm above the previous profile so that overlap is reduced and the evolution of surface features can be seen.   Surface features like ripple crests can be tracked over a number of successive profiles and the slopes of imaginary lines connecting these features are an indication of their horizontal speed away from the belt surface.  It is clear that close to the belt surface, the ripple features last for only a few, say about five, profiles and their paths have a high slope relative to horizontal, indicating relatively slow motion away from the belt.  This region appears to be more chaotic than the region to the right.  In this outer region, surface features remain visible over many frames giving support to the idea that they are freely propagating waves.  The red line in the figure, which was drawn by eye to approximate that slope of the imaginary lines connection the ripple crests in the outer region, corresponds to a velocity of about 34~cm/s.  It  should be kept in mind that this is only the $y$-component of the phase speed.

\begin{figure*}[p!]
\begin{center}
\begin{tabular}{cc}
\includegraphics[trim=0 0.0in 0 0.00in,clip=true,width=2.75in]{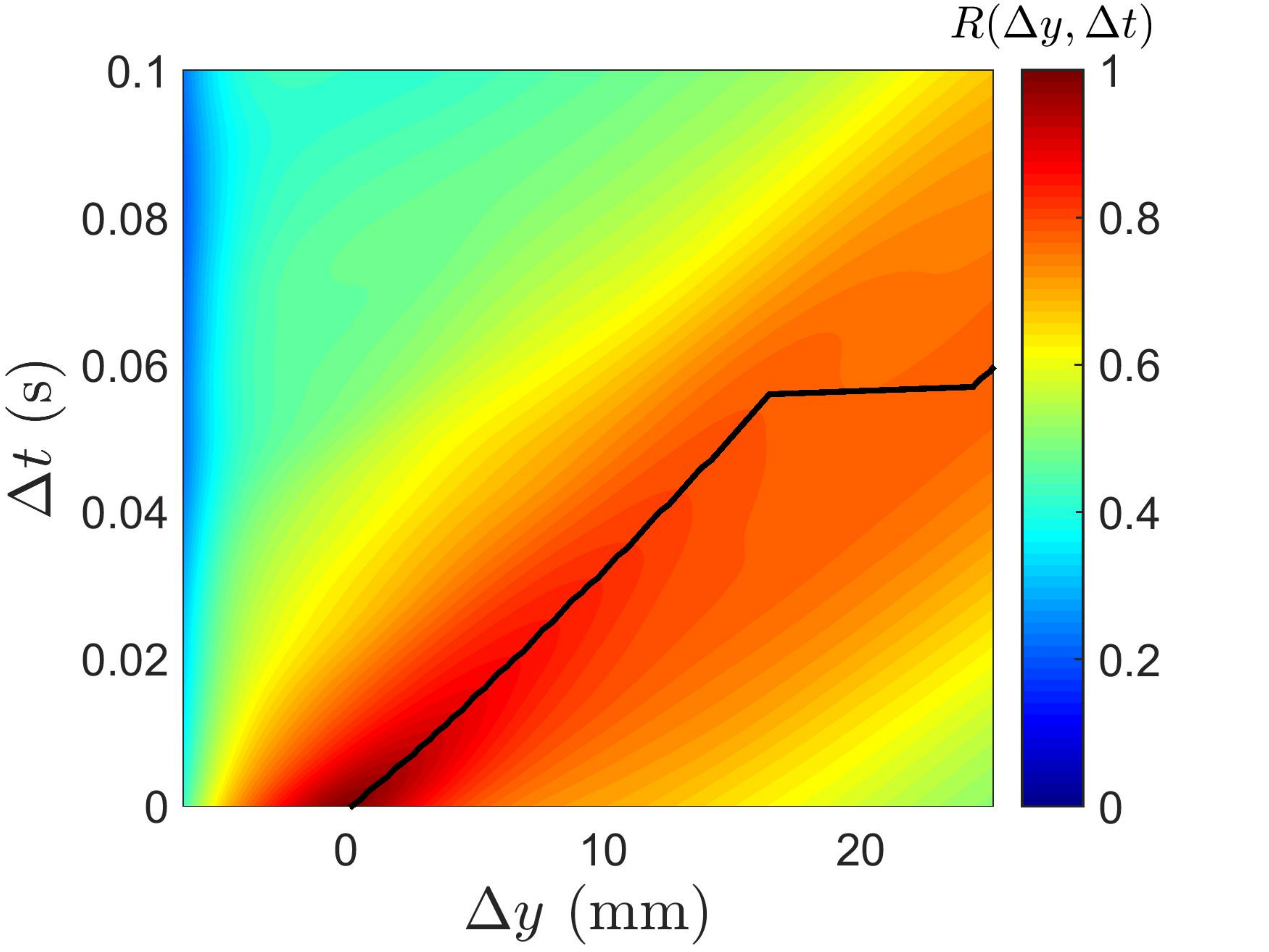} &
\includegraphics[trim=0 0.0in 0 0.00in,clip=true,width=2.75in]{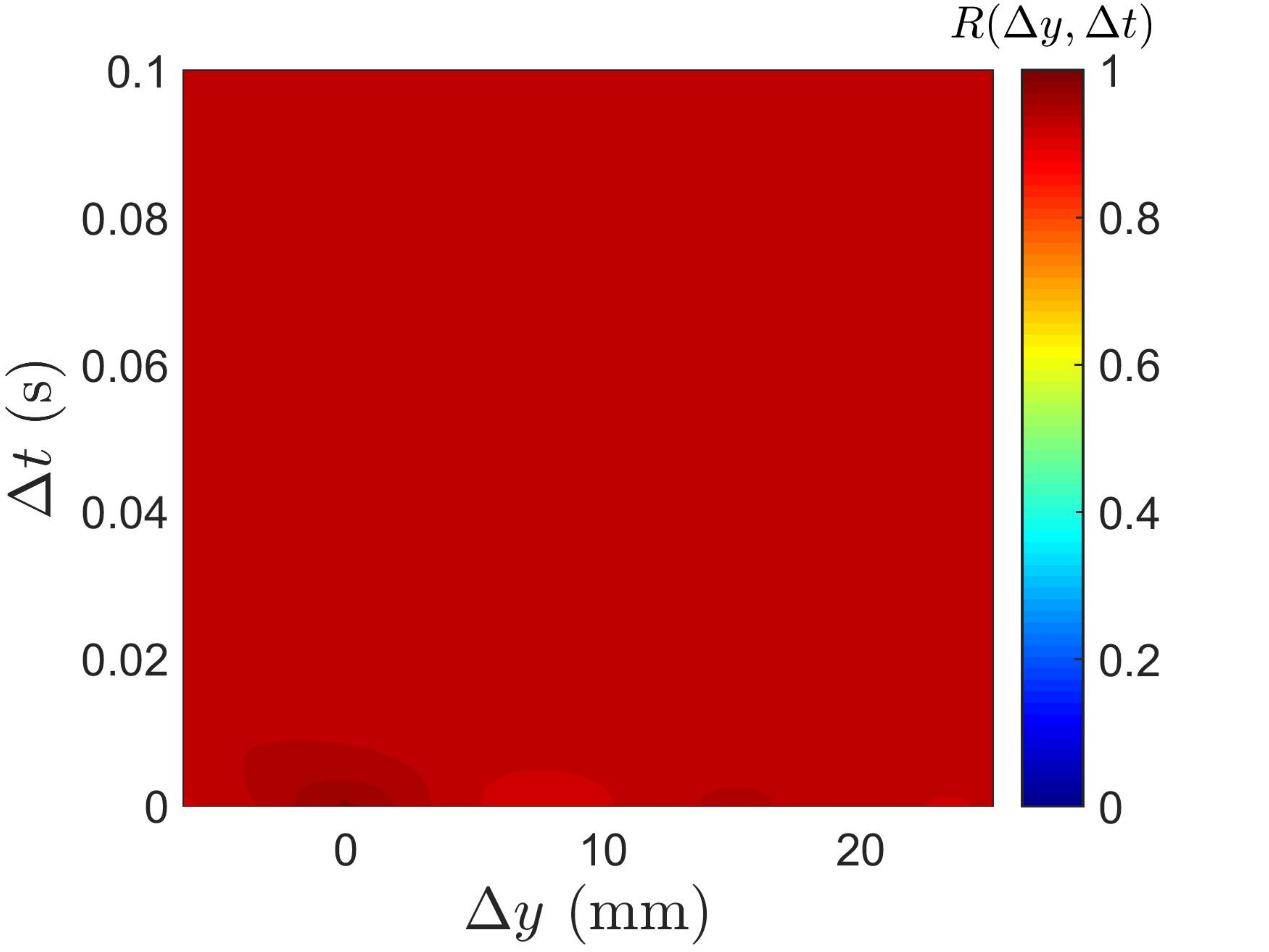}\\
(\textit{a}) & (\textit{b}) \\
\includegraphics[trim=0 0.0in 0 0.00in,clip=true,width=2.75in]{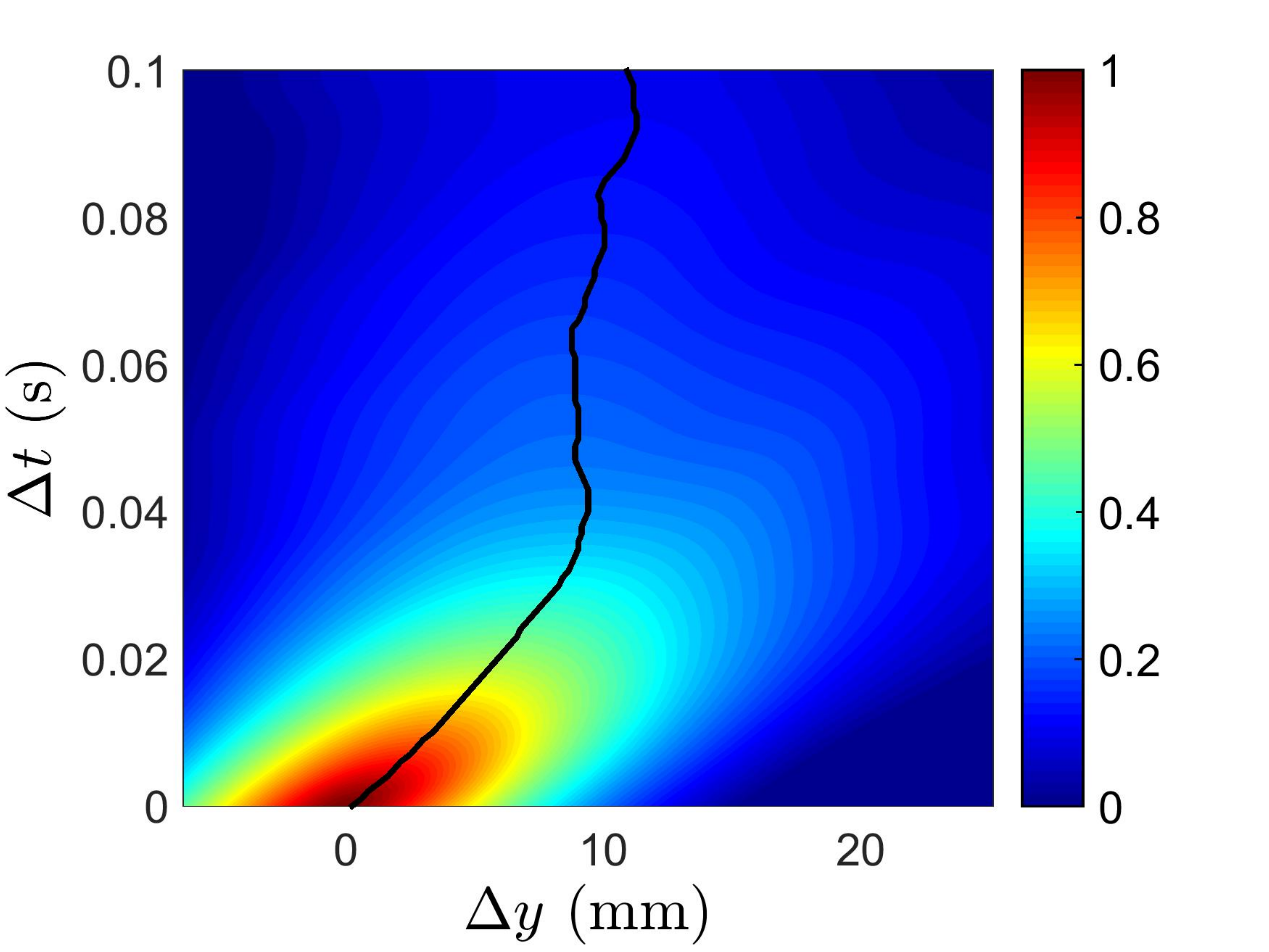} &
\includegraphics[trim=0 0.0in 0 0.00in,clip=true,width=2.75in]{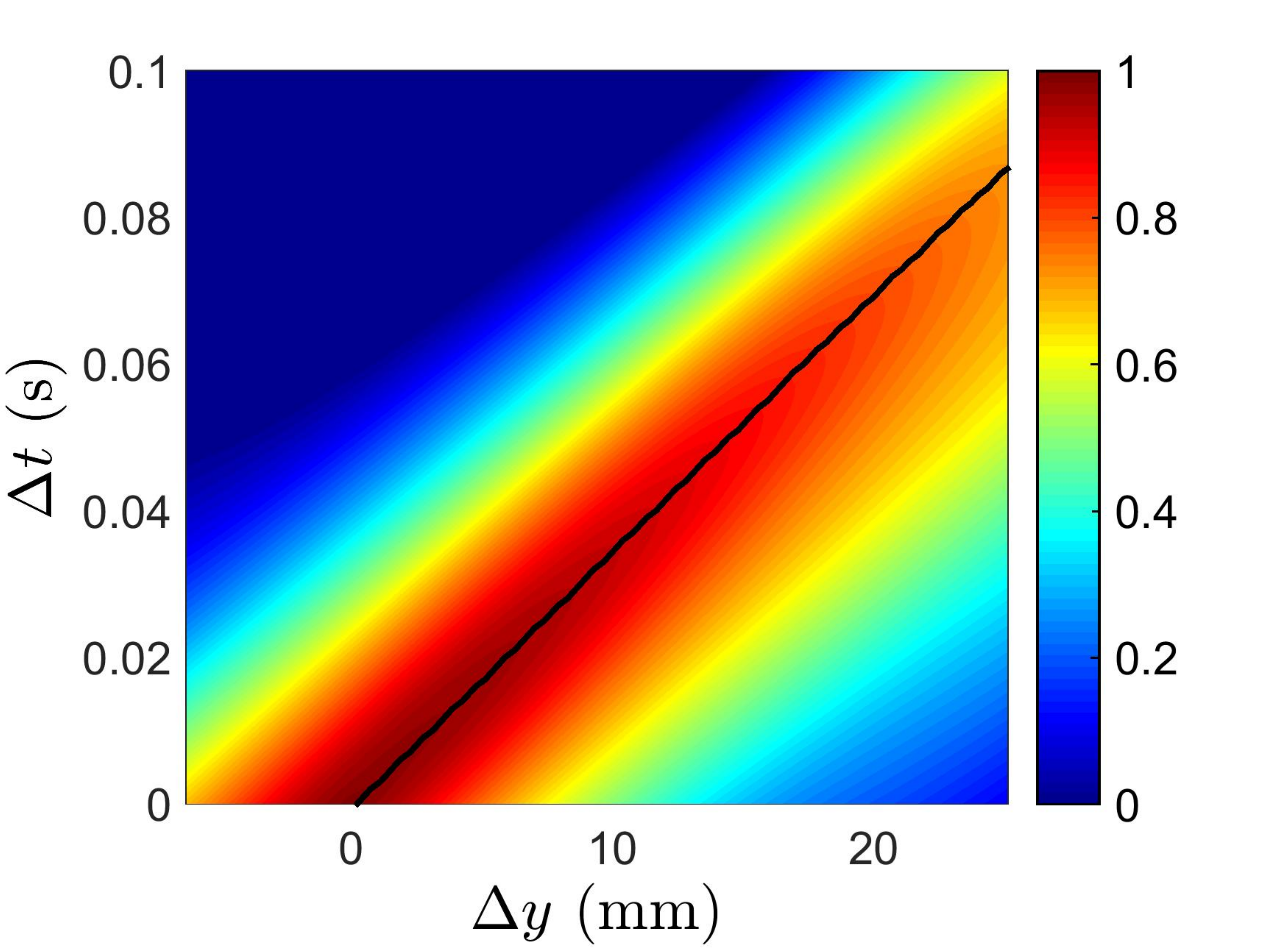}\\
(\textit{c}) & (\textit{d}) \\
\includegraphics[trim=0 0.0in 0 0.00in,clip=true,width=2.75in]{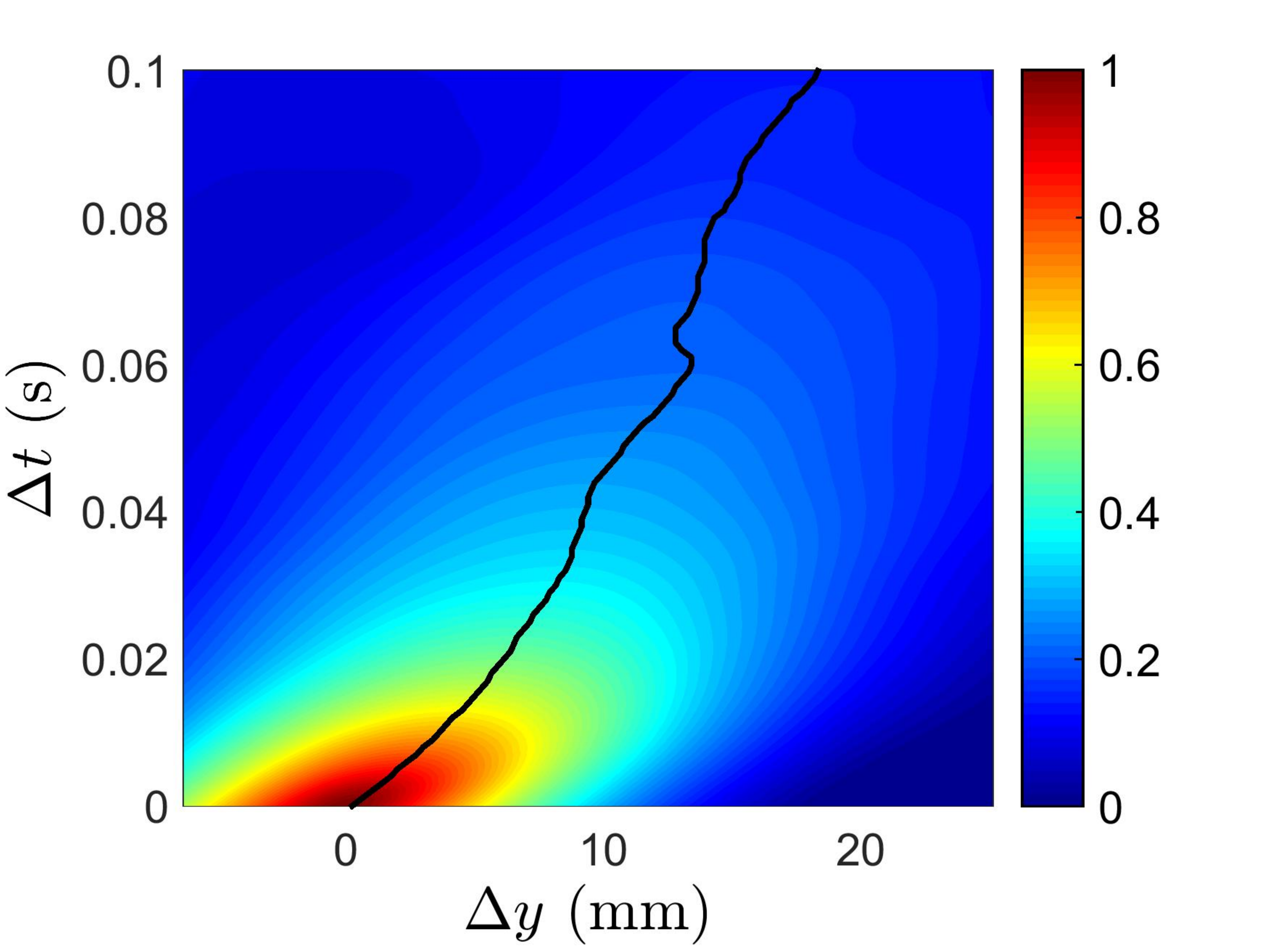} &
\includegraphics[trim=0 0.0in 0 0.00in,clip=true,width=2.75in]{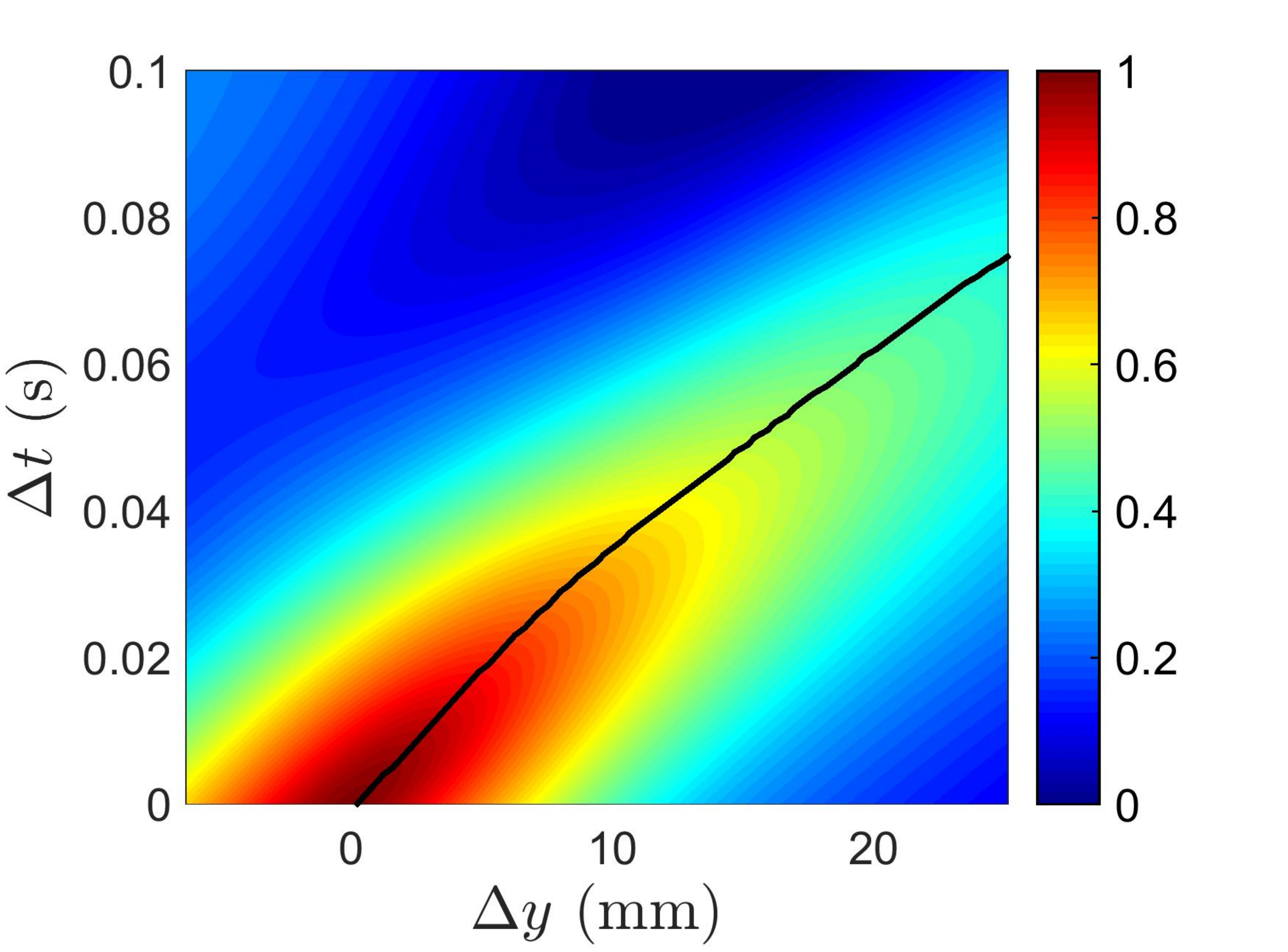}\\
(\textit{e}) & (\textit{f})\\
\end{tabular}
\end{center}
\vspace*{-0.2in} \caption{Cross correlation maps for profiles from a belt launch to 3~m/s. The left column shows cross correlation maps averaged over a region close to the belt (0.6~cm $> y >$ 6.9~cm) and those in the right column are averaged over a region far from the belt (21.4~cm $> y >$ 27.6~cm). The three rows from top to bottom contain plots from $x$ values of 0 to 5~m, 5.85 to 17.85~m, and 17.85 to 29.85~m, respectively. The black line in each plot indicates the $\Delta x$ location of the maximum correlation for each $\Delta t$ slice.}\label{fig:crosscorr}
\end{figure*} 

In order to study the differences in the propagation of free surface ripples in a more quantitative way, we can perform cross correlation between profiles both close to the belt (0.6~cm $> y >$ 6.9~cm) and far away (21.4~cm $> y >$ 27.6~cm), as shown in Figure~\ref{fig:crosscorr}. The cross correlation function $R$ is defined as:
\[\scalebox{1}{$
R= \frac{\sum(Z_1(y,t)-\bar{Z_1})(Z_2(y+\Delta y, t+\Delta t)-\bar{Z_2})}{\sqrt{\sum(Z_1(y,t)-\bar{Z_1})^2}\sqrt{\sum(Z_2(y+\Delta y, t+\Delta t)-\bar{Z_2})^2}},$}
\]
where $y$ is the horizontal wall-normal coordinate, $t$ is the time at which each profile is measured, $\Delta y$ and $\Delta t$ are the spatial and temporal shifts between correlated profiles, and $Z_1$ and $Z_2$ are the two profiles which are being correlated. This cross correlation is performed for all profiles over lengths of belt travel from $x$ = 0 to 30~m, and the resulting correlation maps are summed into three groups from $x$ = 0 to 5~m, 5.85~m to 17.85~m, and 17.85~m to 29.85~m, which are considered to be early run, mid run, and late run, respectively. This procedure is repeated for belt speeds of 3, 4, and 5~m/s. According to classical correlations in \citet{Schlichting}, the boundary layer thickness at 17.85~m (the boundary between mid run and late run) is approximately 17.6 to 19.6~cm, depending on belt speed. This would place the far field interrogation region outside of the influence of the boundary layer in the beginning and middle of the run. The boundary layer reaches the left edge of this far region after 20 to 22.7~m of belt travel, which would indicate that late in the run, the boundary layer may begin to influence the free surface effects in the interrogation region far from the belt.

\begin{figure}
\begin{center}
\includegraphics[trim=0.0in 0.0in 0 0.00in,clip=true,width = 3in]{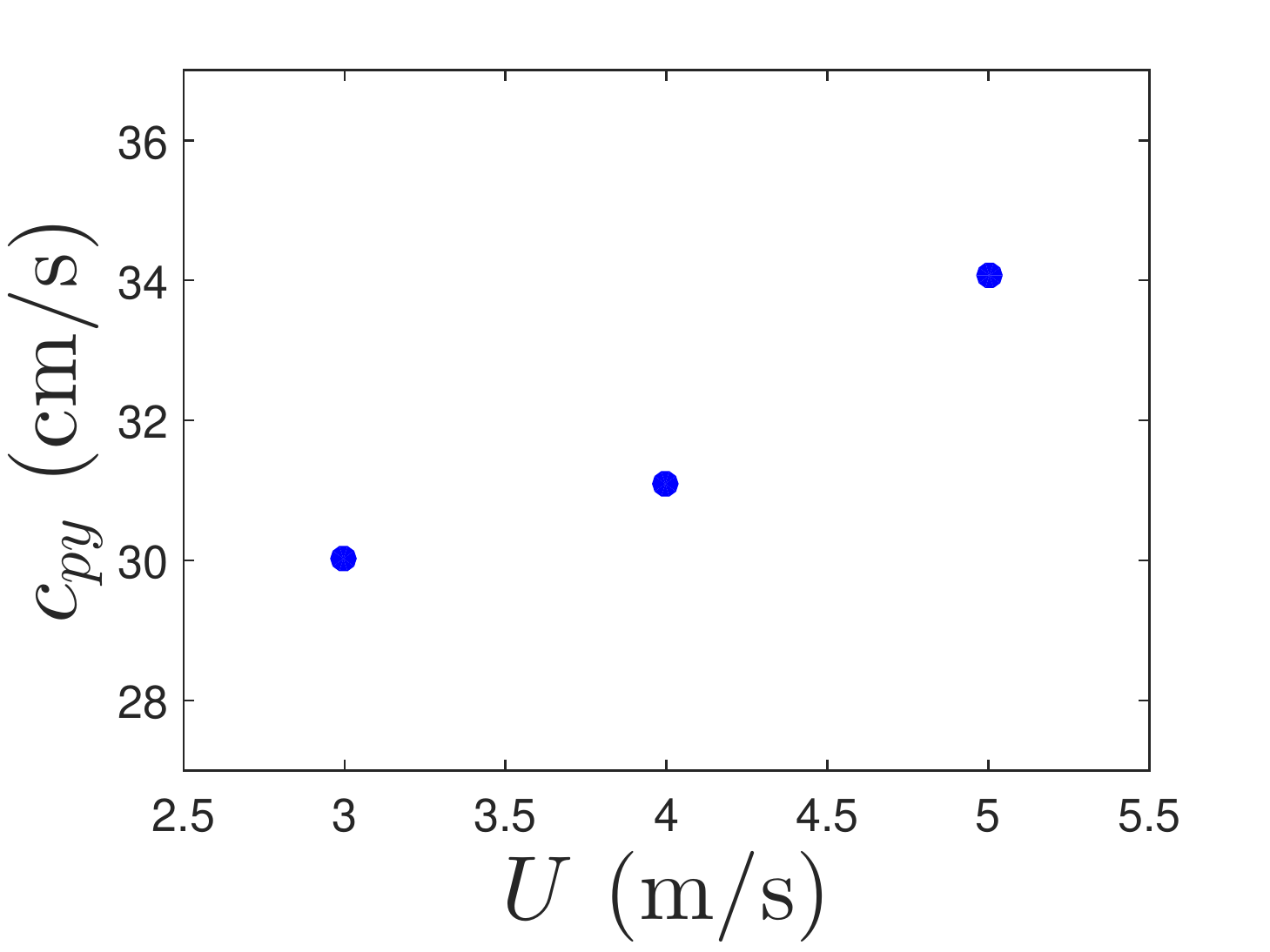} \\
\end{center}
\vspace*{-0.1in} \caption{A plot of the $y$ component of phase speed versus belt speed for the middle portion of each run far from the belt.} \label{fig:phase_speed}
\end{figure}

\begin{figure*}[p!]
\begin{center}
\begin{tabular}{cc}
\includegraphics[trim=0.5in 0.0in 0in 0.00in,clip=false,width=2.55in]{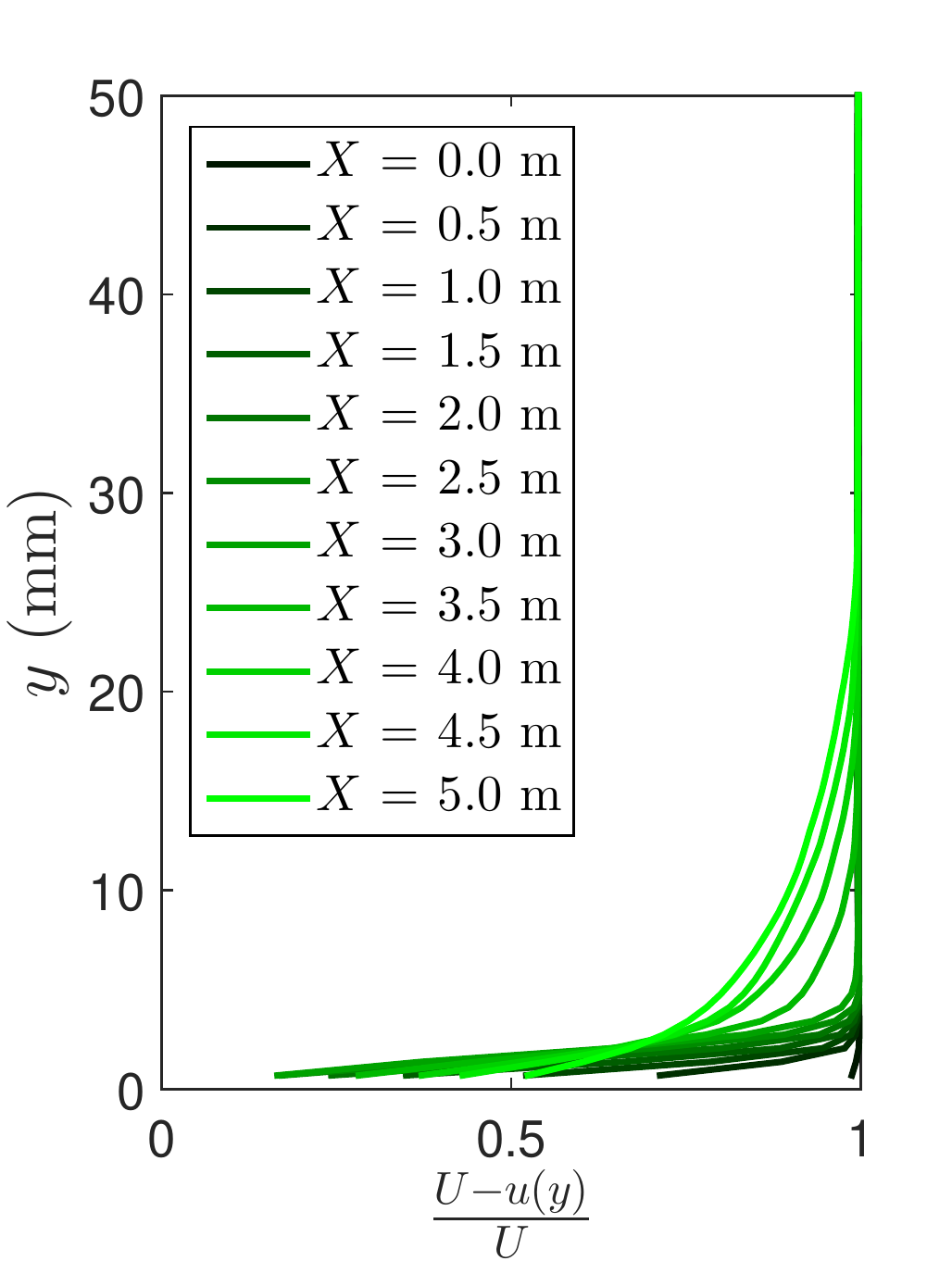} \hspace{.25in} &
\includegraphics[trim=0.5in 0.0in 0in 0.00in,clip=false,width=2.55in]{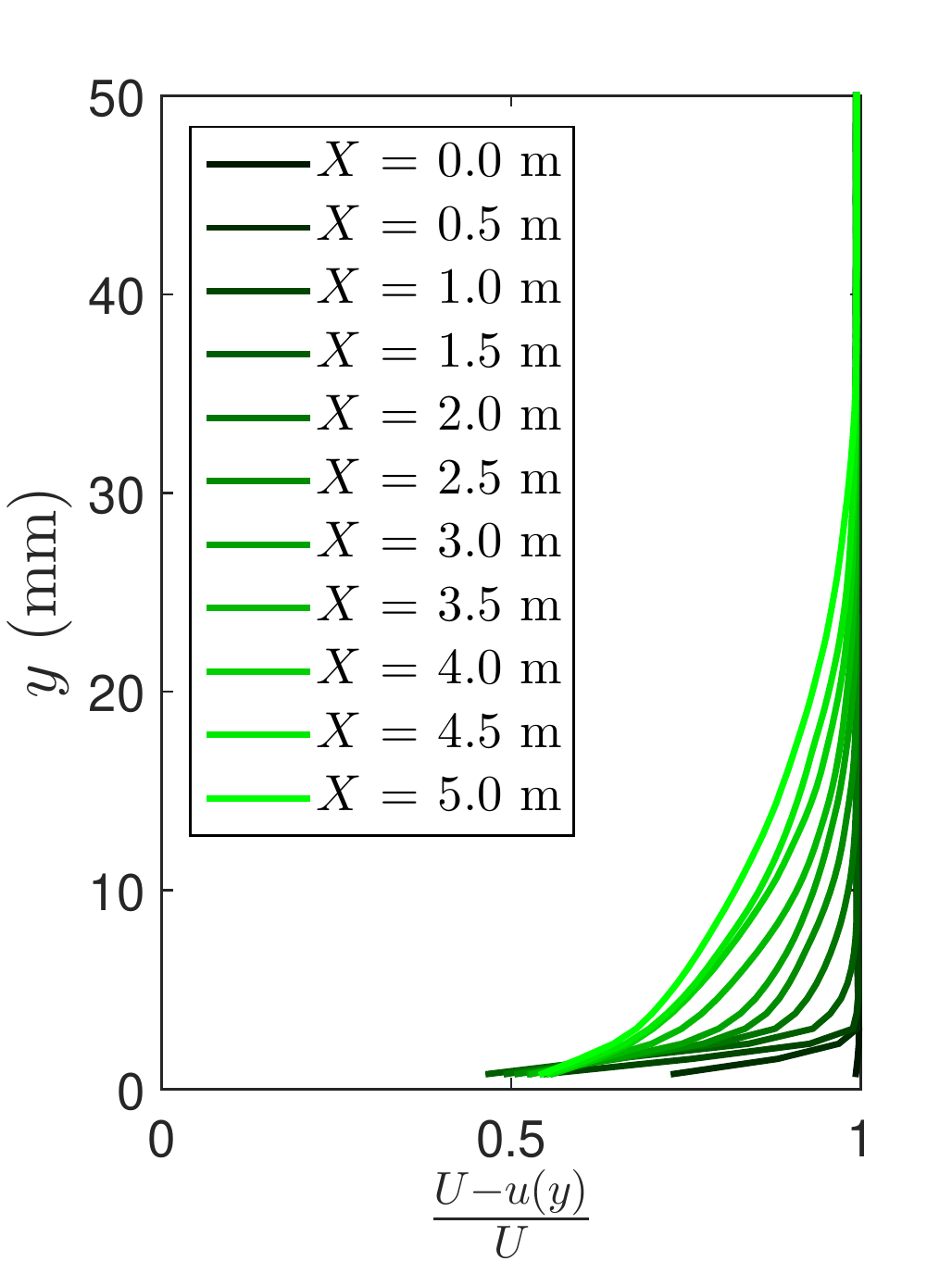}\\
(\textit{a}) & (\textit{b})\\
\includegraphics[trim=0.5in 0.0in 0in 0.00in,clip=false,width=2.55in]{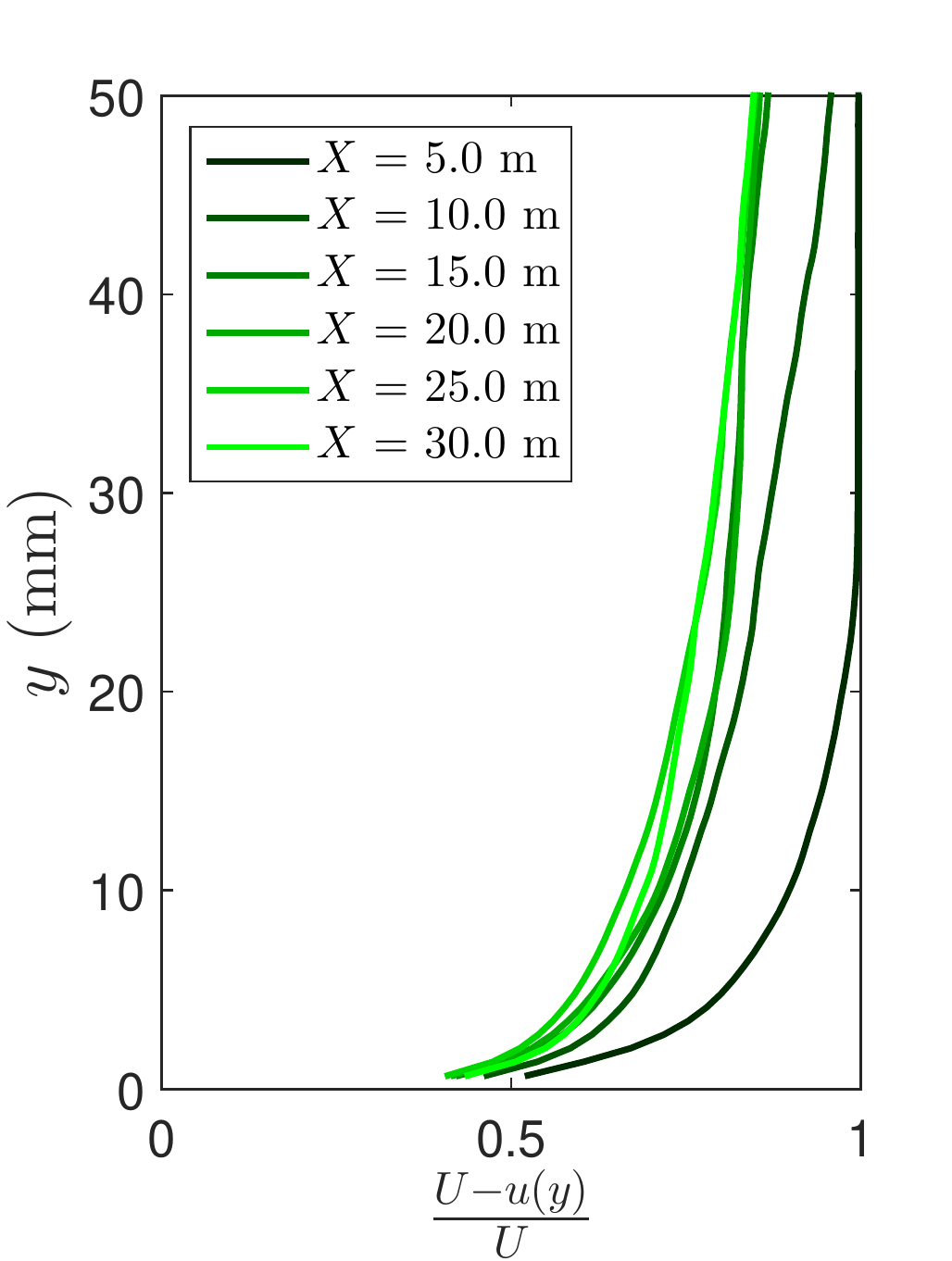} \hspace{.25in} &
\includegraphics[trim=0.5in 0.0in 0in 0.00in,clip=false,width=2.55in]{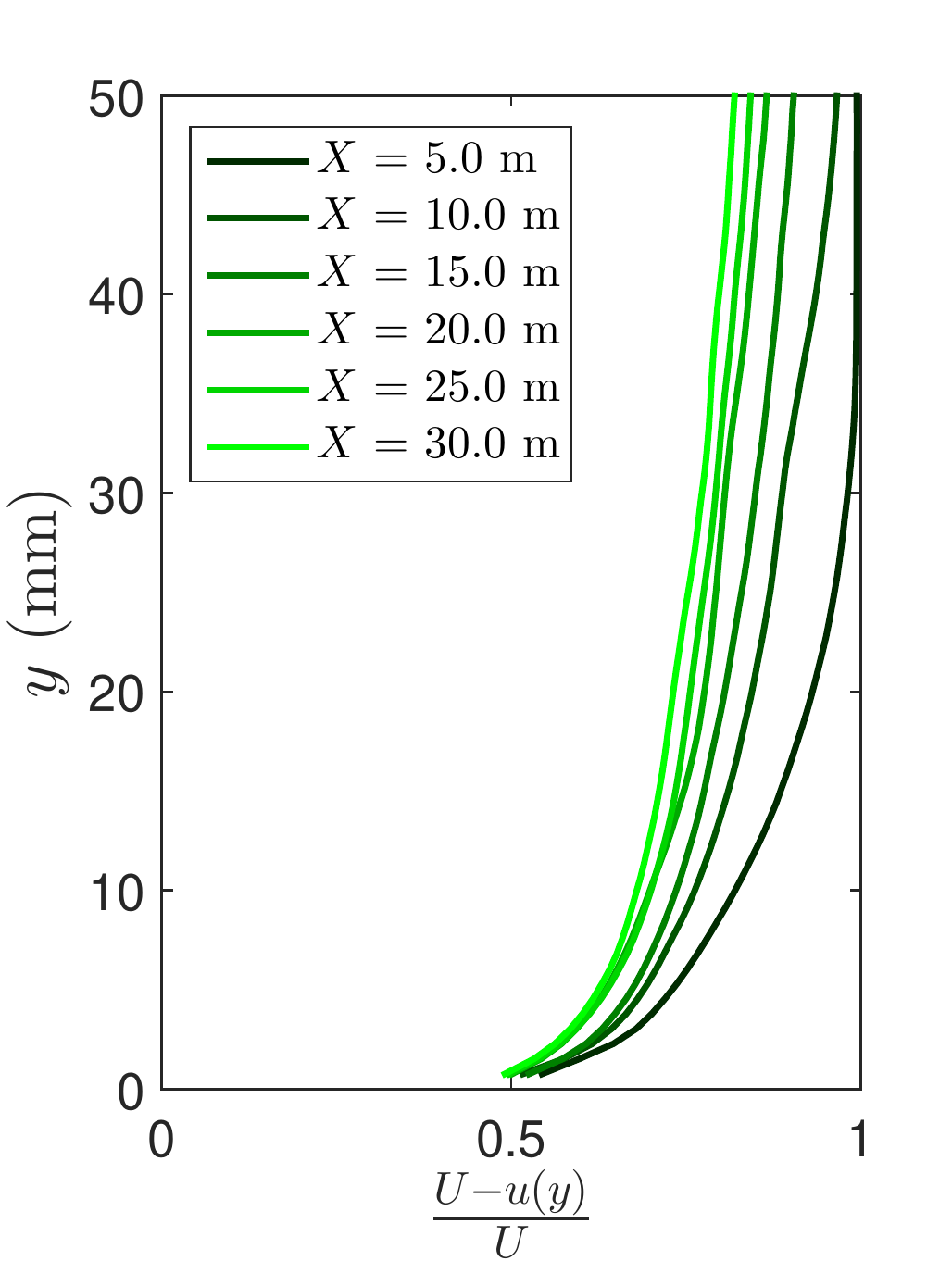}\\
(\textit{c}) &(\textit{d})
\end{tabular}
\end{center}
\vspace*{-0.2in} \caption{Mean streamwise velocity profiles at a belt speed of $U$ = 3~m/s for (\textit{a}), (\textit{c}) $D$ = 14~cm and (\textit{b}), (\textit{d}) $D$ = 2.5~cm. Images (\textit{a}) and (\textit{b}) each plot a profile at each 0.5~m of belt travel from $x$ = 0 to 5~m, while (\textit{c}) and (\textit{d}) each plot a profile at each 5~m of belt travel from $x$ = 5 to 30~m.} \label{fig:V3_mean_profiles}
\end{figure*}

The set of cross correlation maps in Figure~\ref{fig:crosscorr} contains plots from a belt speed of 3~m/s. The left column shows these maps from close to the belt and the right column shows maps far from the belt. The first row is from early in the run, the second row is from the middle of the run, and the third row is from late in the run. Early in the run, the cross correlation map near the belt, shown in (\textit{a}) shows some level of coherence, with a fairly high correlation value sustained for relatively long $\Delta t$, but the correlation peak is wide, indicating that no single dominant velocity exists among the ripples in this region. In the mid and late run portion of the launch, the cross correlation maps close to the belt, shown in (\textit{c}) and (\textit{e}), appear to show a very short correlation time, with free surface features essentially becoming uncorrelated after $\Delta t$ of about 0.4~s. These maps also do not appear to show a single dominant propagation velocity, confirming the idea that these near-belt ripples change rapidly and do not appear to propagate freely. Far from the belt, the cross correlation map early in the run in (\textit{b}) is flat because no free surface ripples have reached that location yet and the surface is flat. In the middle of the run, far from the belt, as shown in (\textit{d}), the cross correlation map appears to form a clearly defined ridge, which retains a high cross correlation value for relatively long $\Delta t$ and appears to show a single dominant velocity, which appears to indicate that the free surface ripples may become freely propagating outside of the influence of the boundary layer. Late in the run, this far field region shows a weakening of the cross correlation ridge, lending credence to the idea that the boundary layer begins to influence this interrogation region later in the run.

This analysis is repeated for all three belt speeds with similar results. One notable difference between these cross correlation maps is that the clearly defined ridge far from the belt in the middle of each run, as seen in Figure~\ref{fig:crosscorr}-(\textit{d}), appears to have a different slope depending on belt speed, shown in Figure~\ref{fig:phase_speed}. This plot shows a clear increase of wall-normal propagation speed with belt speed. Assuming that these are freely propagating waves that follow the dispersion relation for linear gravity waves in deep water, these phase speeds would correspond to wavelengths of 5.78, 6.19, and 7.44~cm, respectively. Because these recorded waves only consist of the $y$-component of each wave, further information of the $x$-direction wave component is necessary to determine whether these waves are truly propagating with a higher speed or if the propagation diirection is changing.

\subsection{Velocity Profiles}

\begin{figure}
\begin{center}
\includegraphics[trim=0.0in 0.0in 0 0.00in,clip=true,width = 3in]{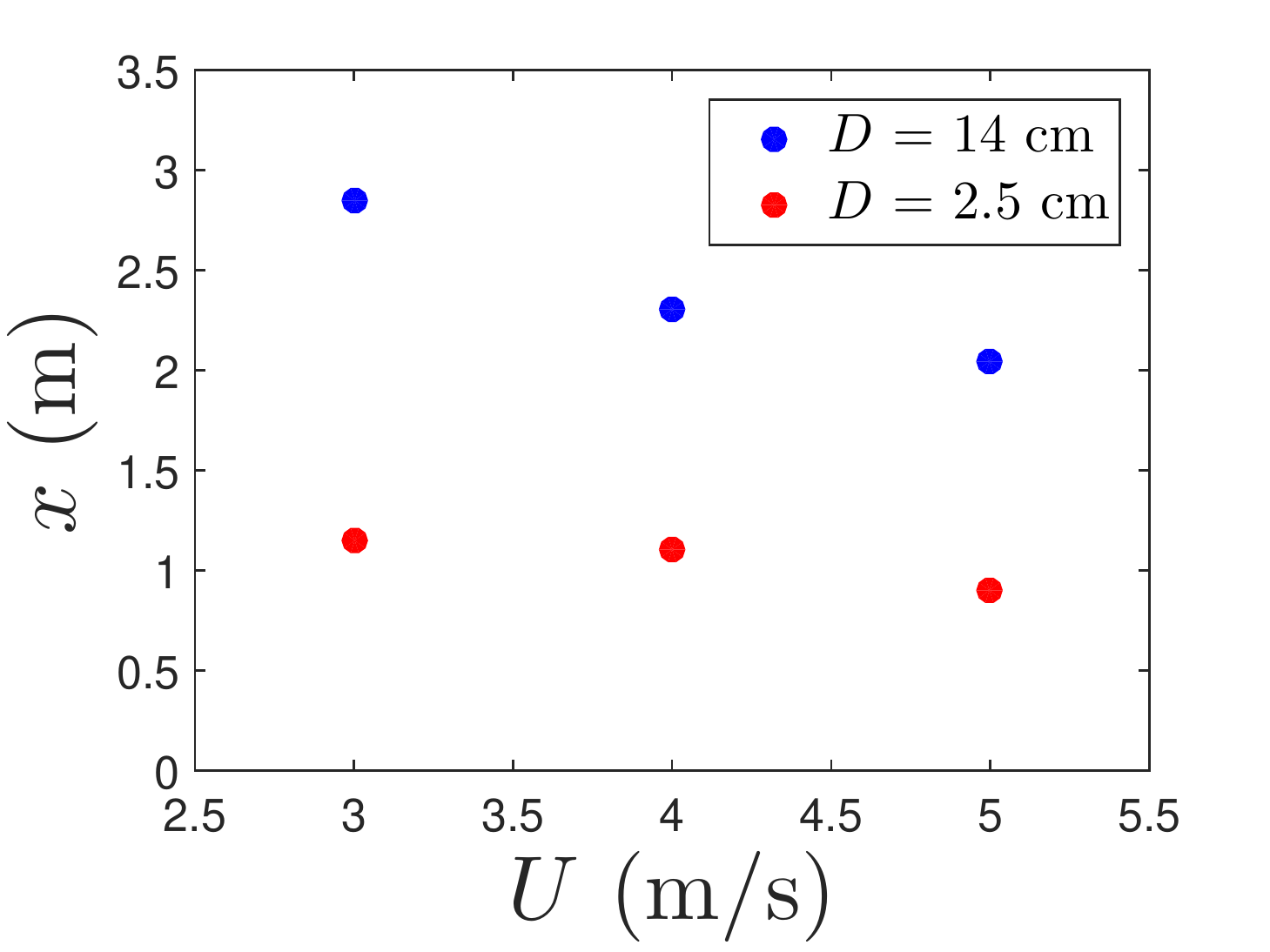} \\
\end{center}
\vspace*{-0.1in} \caption{A plot of the $x$ location at the onset of rapid boundary layer growth versus belt speed.} \label{fig:mean_burst}
\end{figure}

Velocity vector fields are recorded in both horizontal and vertical planes in separate experiments. Because the boundary layer evolves temporally along the entire belt at the same time, the processed PIV vector fields were able to be averaged in the streamwise direction for greater statistical convergence. Experiments at each light sheet location and belt speed were also repeated 20 times for the purpose of ensemble averaging. In addition, some averaging in $x$ by $\pm$ 8 frames was performed for each condition, which corresponds to 9.6~cm, 12.8~cm, and 16~cm of belt travel for belt speeds of 3, 4, and 5~m/s, respectively. Velocity profiles, $u(y)$, averaged in this way for a belt speed of 3~m/s can be seen in Figure~\ref{fig:V3_mean_profiles}. The plots in the left column contain profiles from the $D$ = 14~cm condition, while the plots in the right-hand column contain profiles from the $D$ = 2.5~cm condition. The plots in the top row depict profiles from each 0.5~m of belt travel from $x$ = 0 to 5~m, while the bottom row contains profiles at each 5~m from $x$ = 5 to 30~m. It can be seen by comparing the plots in the top row that the boundary layer growth appears quite different between these different conditions. Far from the surface, the profiles from $x$ = 0 to 3~m are very closely clustered together before the growth rate suddenly accelerates after approximately 2.85~m of travel. Near the surface, this burst happens much more quickly, with the growth rate accelerating after only about 1.15~m of belt travel. This rapid boundary layer growth appears to decrease with increasing belt speed, as shown in Figure~\ref{fig:mean_burst}.

\begin{figure}
\begin{center}
\includegraphics[trim=0.0in 0.0in 0 0.00in,clip=true,width = 3in]{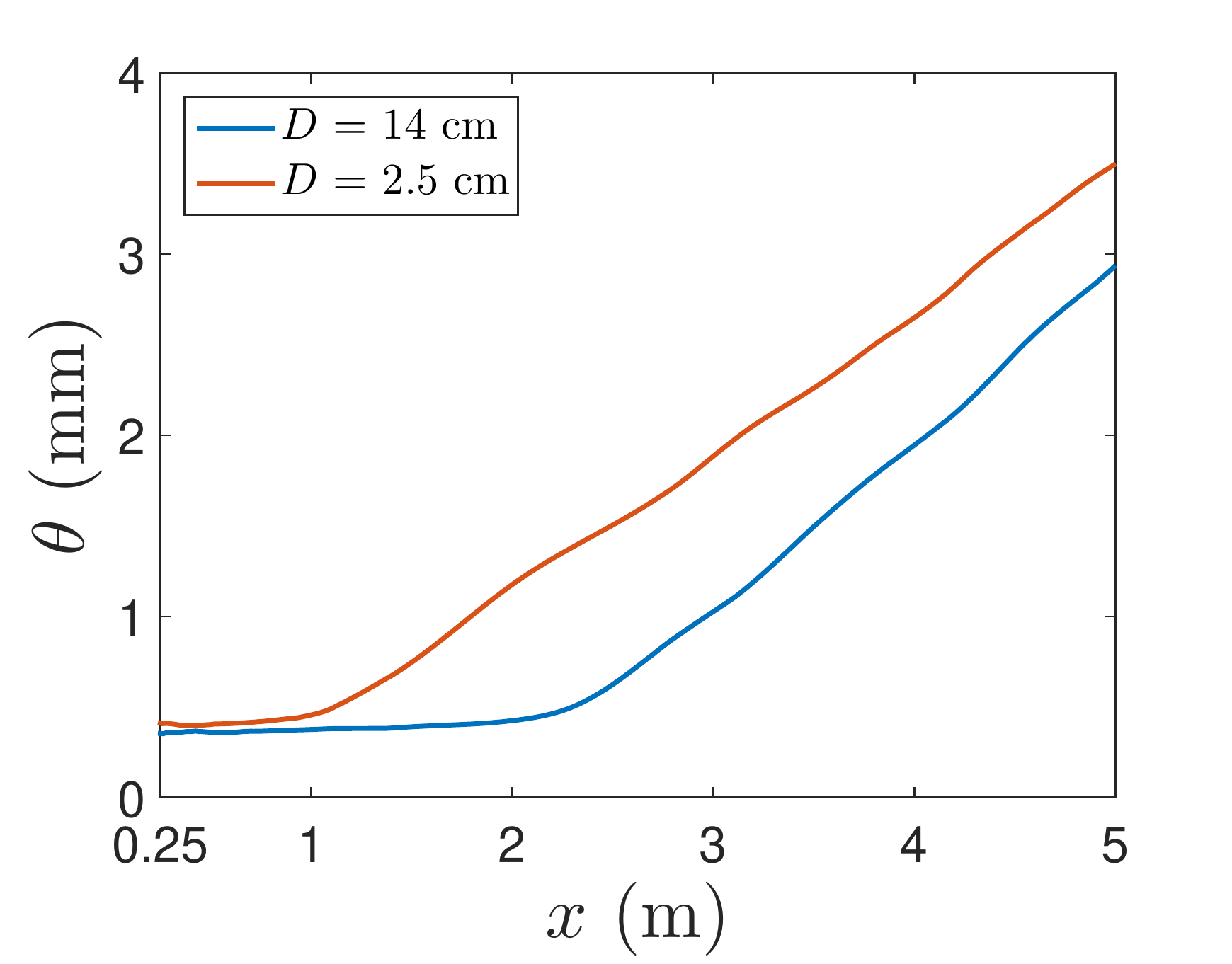} \\
\end{center}
\vspace*{-0.1in} \caption{A plot of the momentum thickness versus $x$ calculated using mean velocity profiles from both horizontal PIV planes with a belt speed of 5~m/s.} \label{fig:momentum_burst}
\end{figure}

The mean velocity profiles can also be used to calculate the momentum thickness in the beginning of each run up until the $x$ location where the boundary layer grows beyond the frame of the measurement region. The momentum thickness is the distance that the surface would have to be displaced in order for a uniform free stream velocity profile to retain the same total momentum and is defined as follows: 
\[
\theta=\int_0^\infty \frac{u(y)}{U}(1- \frac{u(y)}{U})dy.
\]
This relation can be applied to calculate momentum thickness at each $x$ location for both horizontal planes, as shown in Figure~\ref{fig:momentum_burst}. It can be seen in each momentum thickness profile that after an initial period of slow boundary layer growth, a kink in the profile occurs, followed by a higher boundary layer growth rate. This indication of transition to turbulence agrees very well with the $x$ locations of boundary layer growth shown in Figure~\ref{fig:mean_burst}.

\begin{figure*}[htb]
\begin{center}
\begin{tabular}{cc}
\includegraphics[trim=0 0.0in 0 0.00in,clip=true,width=3in]{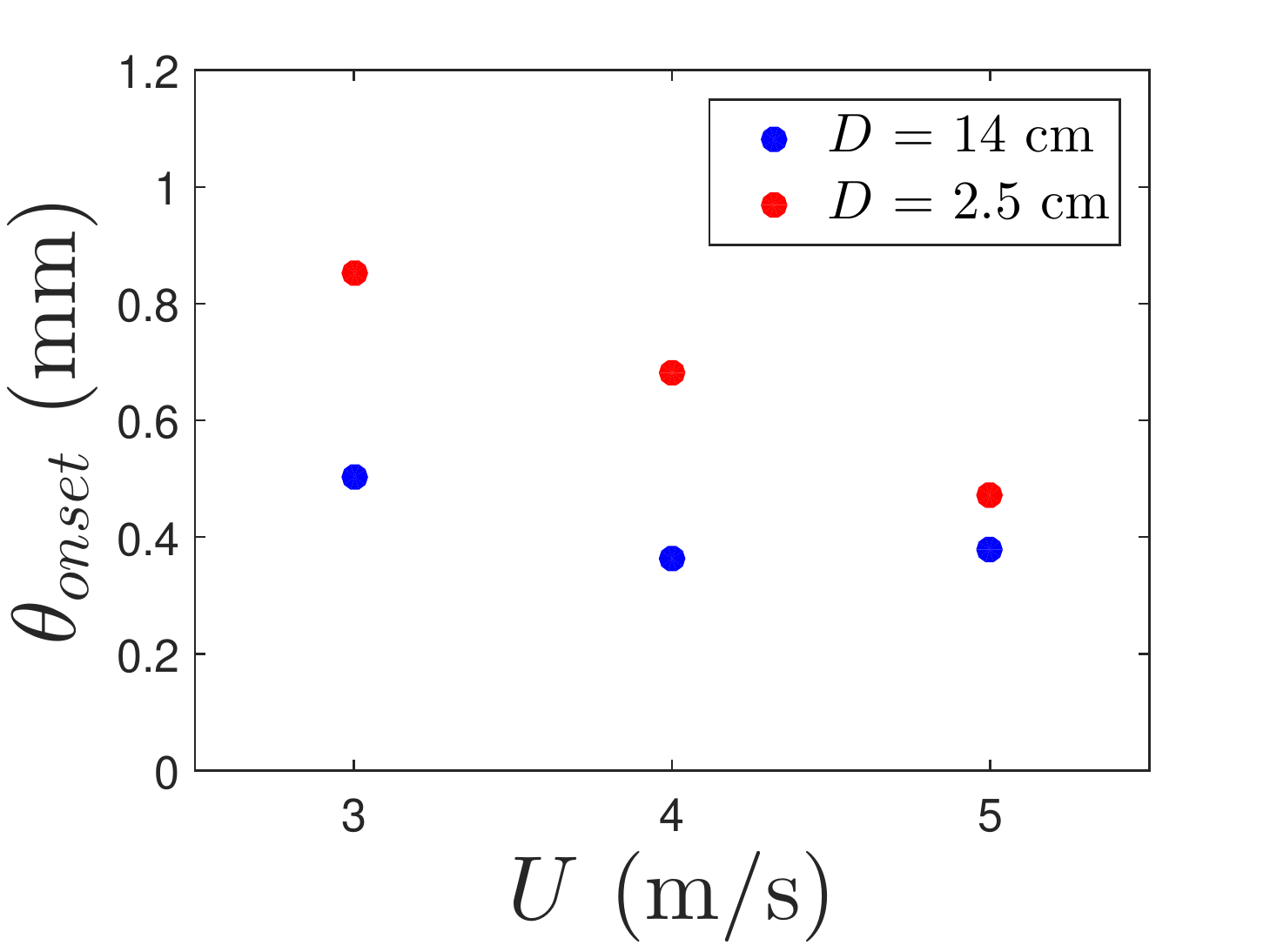}&
\includegraphics[trim=0 0.0in 0 0.00in,clip=true,width=3in]{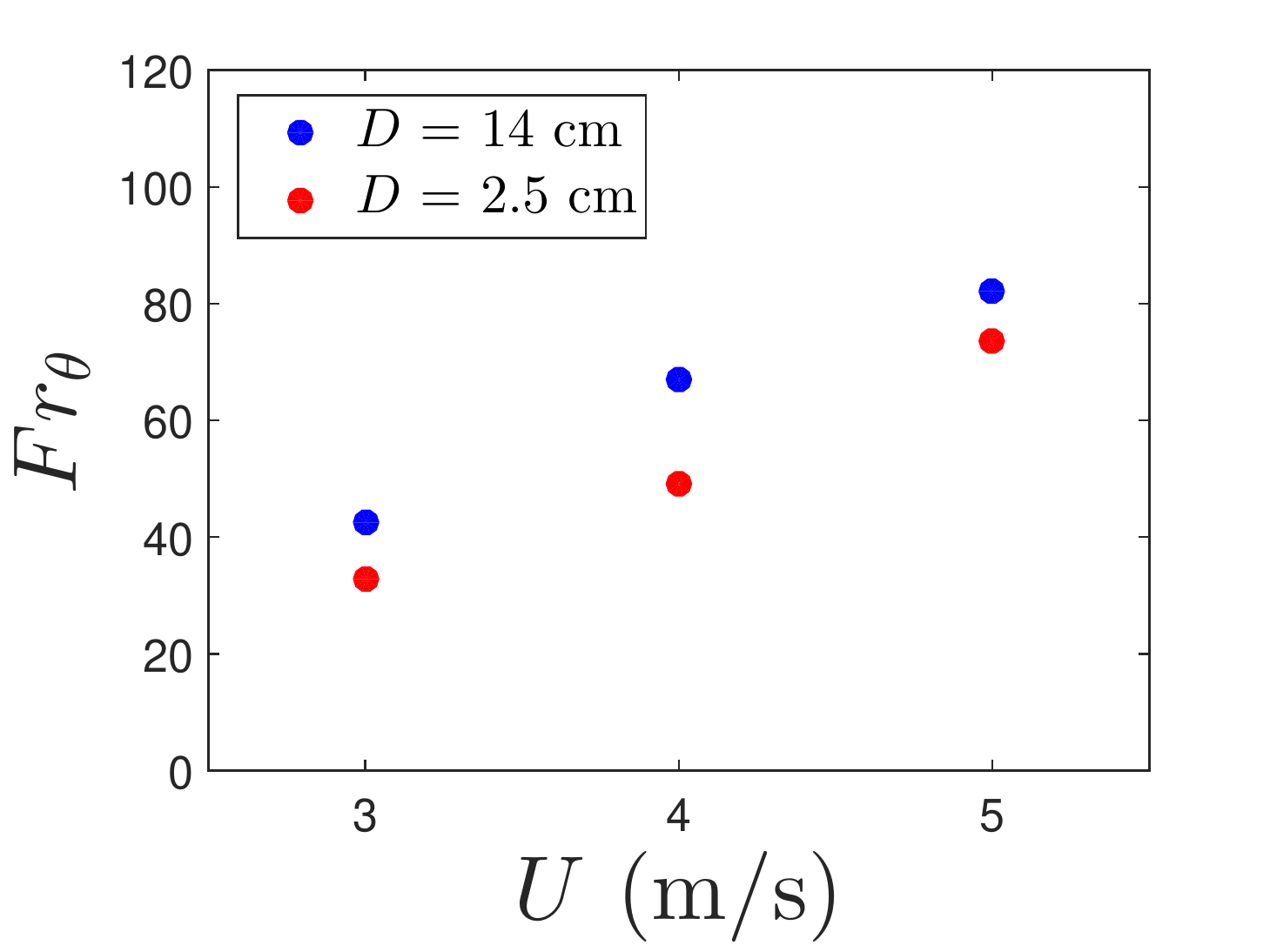}\\
(\textit{a}) & (\textit{b})\\
\includegraphics[trim=0 0.0in 0 0.00in,clip=true,width=3in]{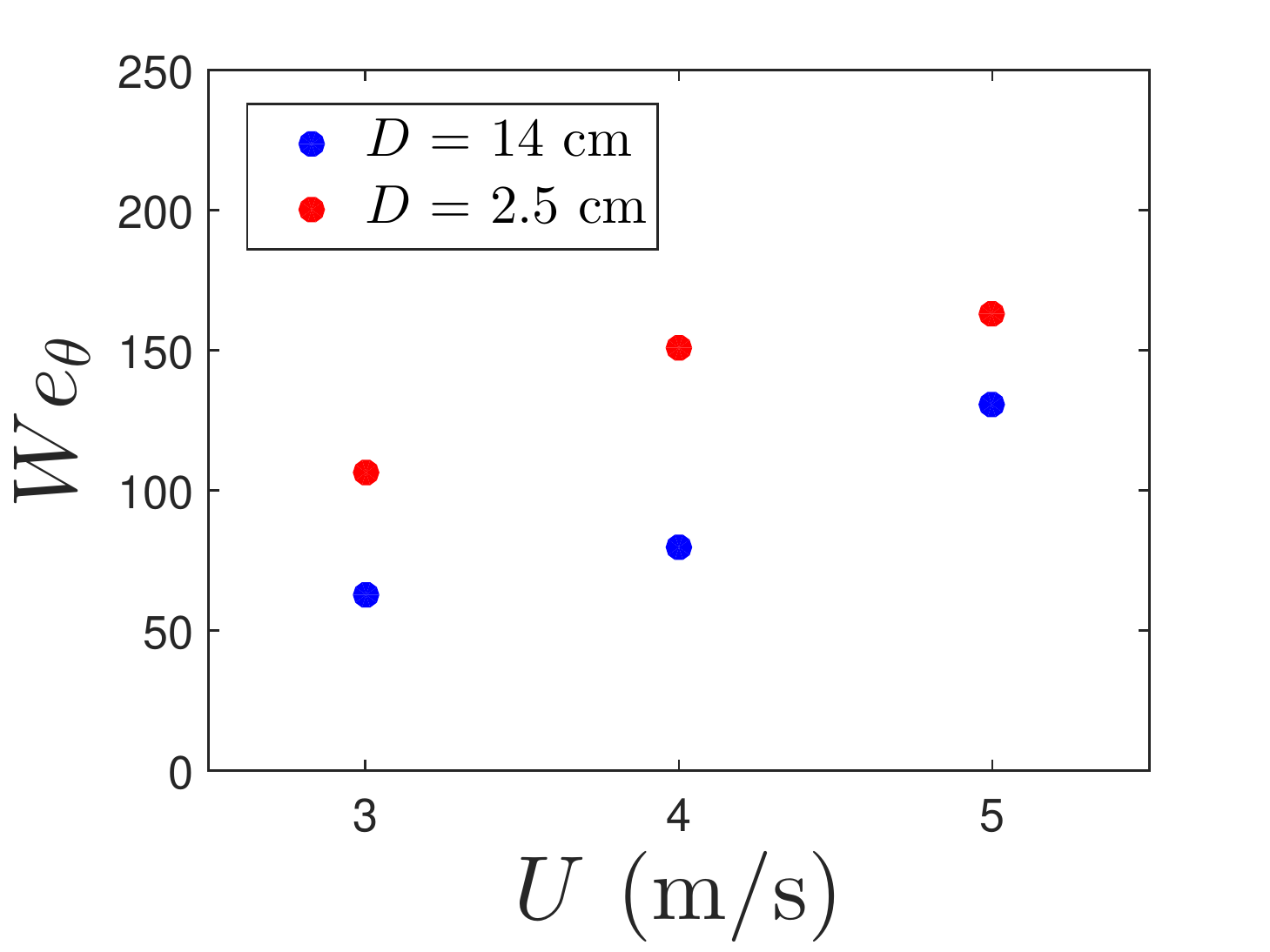}&
\includegraphics[trim=0 0.0in 0 0.00in,clip=true,width=3in]{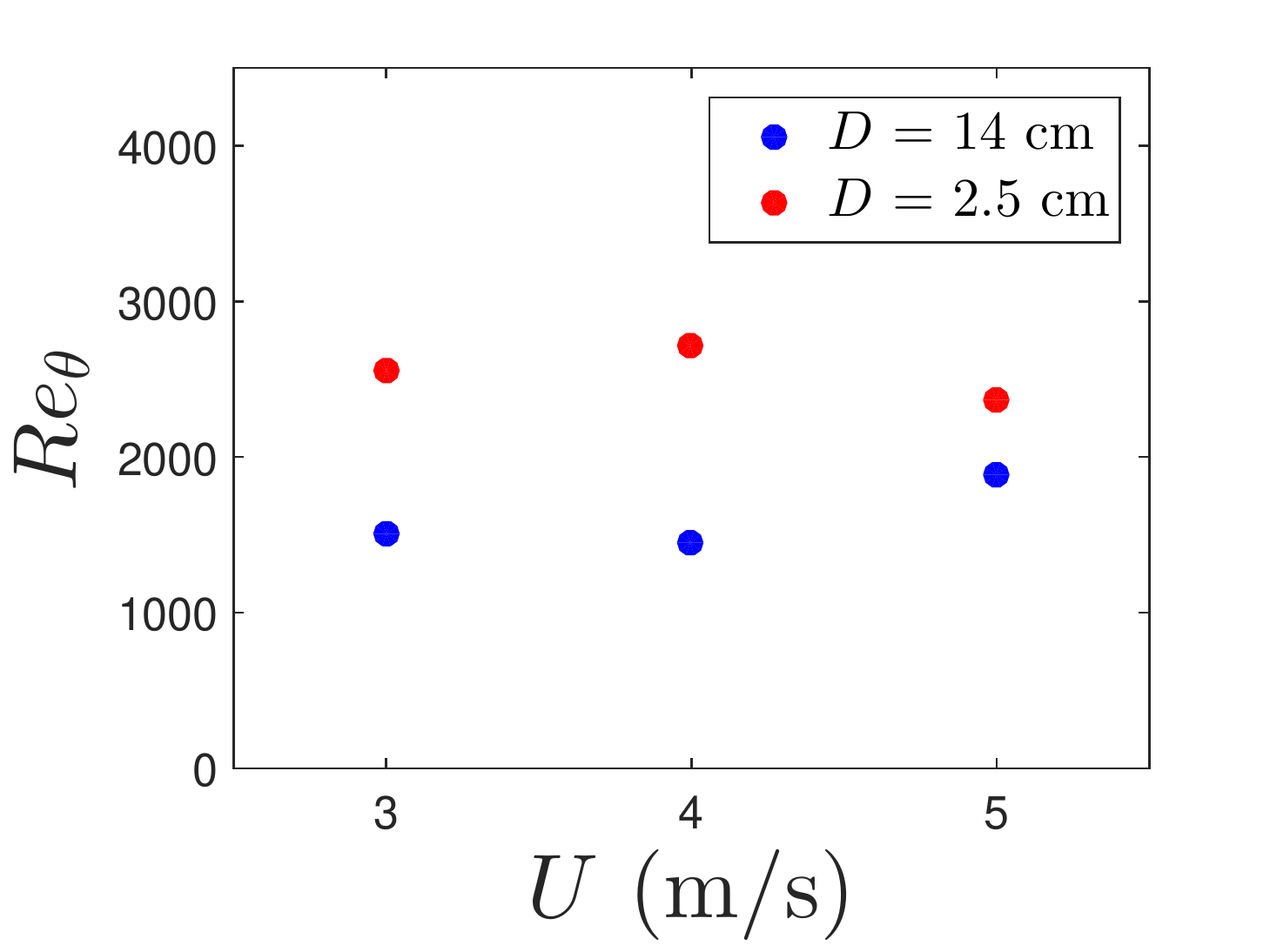}\\
(\textit{c}) & (\textit{d})\\
\end{tabular}
\end{center}
\vspace*{-0.2in} \caption{Plots showing different scaling parameters for bursting onset, calculated using measured velocity profile data. In all plots, blue dots indicate velocity from $D$ = 14~cm and red dots indicate velocity closer to the surface at $D$ = 2.5~cm.} \label{fig:burst_scaling}
\end{figure*}

The values of the boundary layer momentum thickness $\theta$ at the $x$ locations of free surface bursting, see Figure~\ref{fig:bursting}-(\textit{a}), are plotted versus the belt speed $U$ in Figure~\ref{fig:burst_scaling}-(\textit{a}), where the blue dots indicate $\theta$ values obtained using the velocity measurements from the deeper horizontal plane, $D$ = 14~cm, and the red dots indicate values obtained with the velocity measurements from the shallow plane, $D$ = 2.5~cm. In Figure~\ref{fig:burst_scaling}-(\textit{b}), (\textit{c}), and (\textit{d}), the Froude number, Weber number, and Reynolds number, respectively, based on the belt speed $U$ and the momentum thickness at  bursting are plotted versus $U$.   It can be seen in these plots that the momentum thickness is greater closer to the surface, which is in agreement with the idea that the boundary layer grows faster near the free surface, and the momentum thickness at bursting decreases with increasing belt speed. Both the Froude and Weber number at bursting onset appear to show a monotonic increase with belt speed, while the Reynolds number based on momentum thickness appears to provide the most consistent value at bursting onset.

\begin{figure*}[!p]
\begin{center}
\begin{tabular}{cc}
\includegraphics[trim=0.5in 0.0in 0in 0.00in,clip=false,width=2.25in]{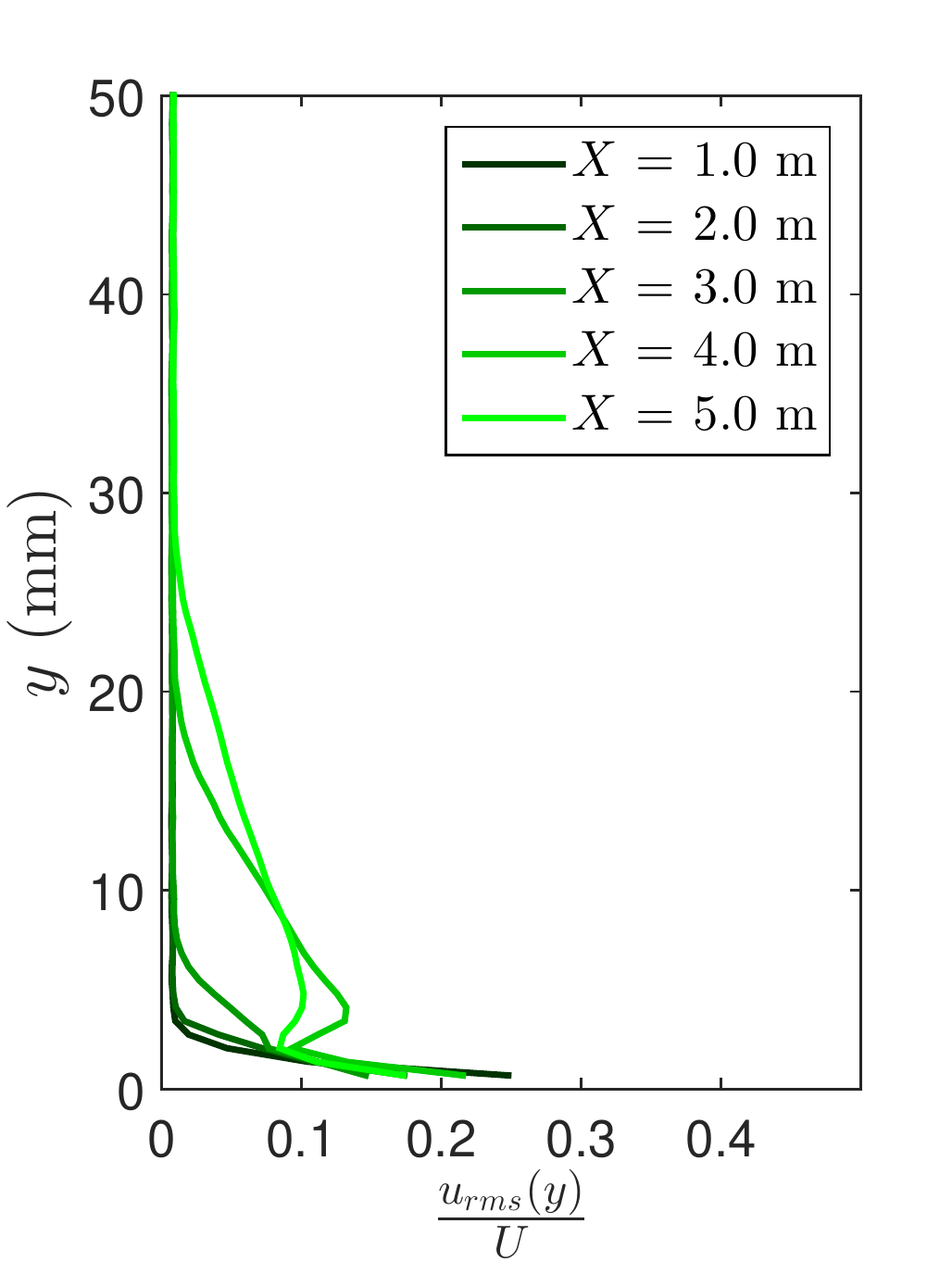} \hspace{.25in} &
\includegraphics[trim=0.5in 0.0in 0in 0.00in,clip=false,width=2.25in]{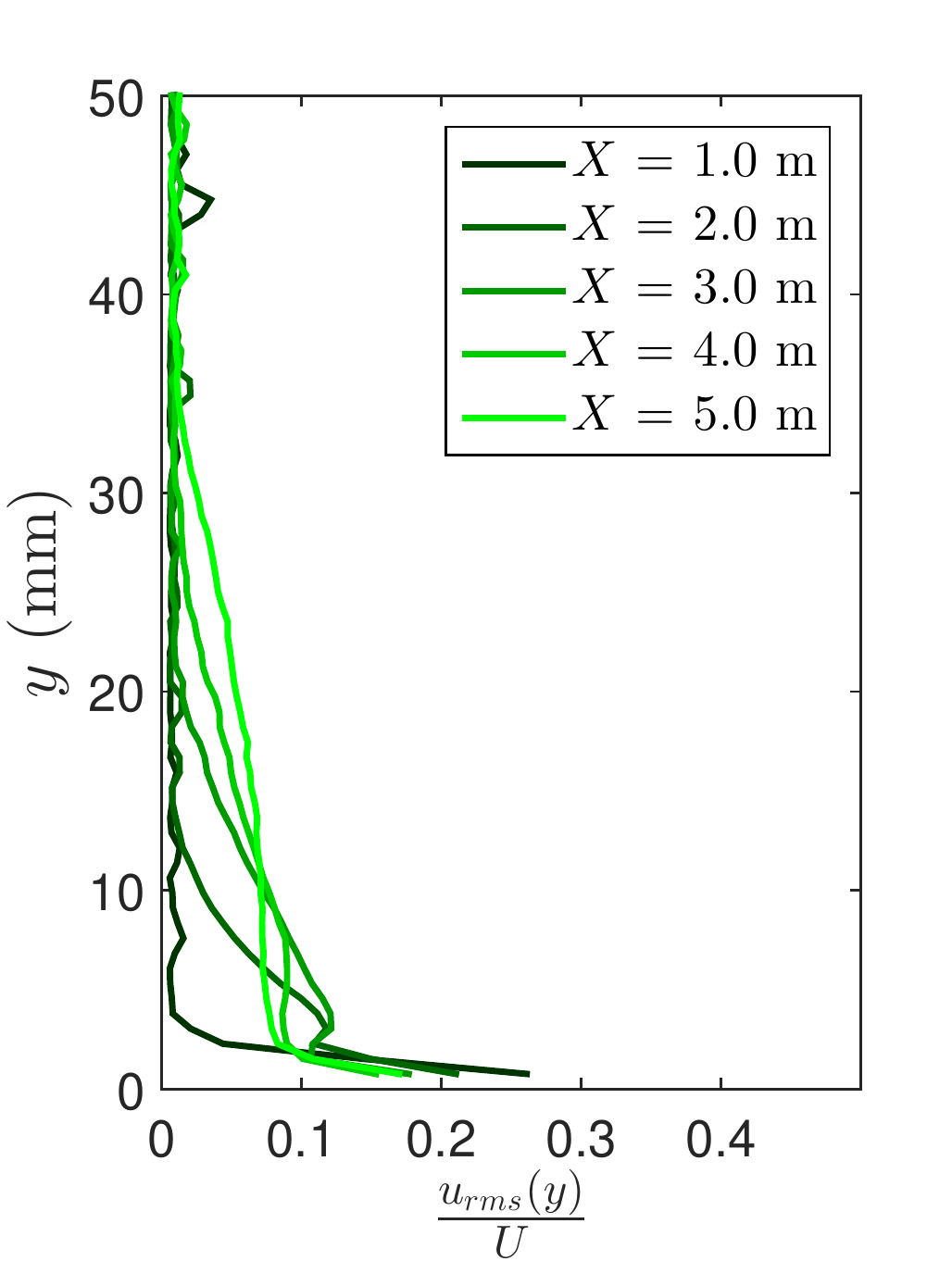}\\
(\textit{a}) & (\textit{b})\\
\includegraphics[trim=0.5in 0.0in 0in 0.00in,clip=false,width=2.25in]{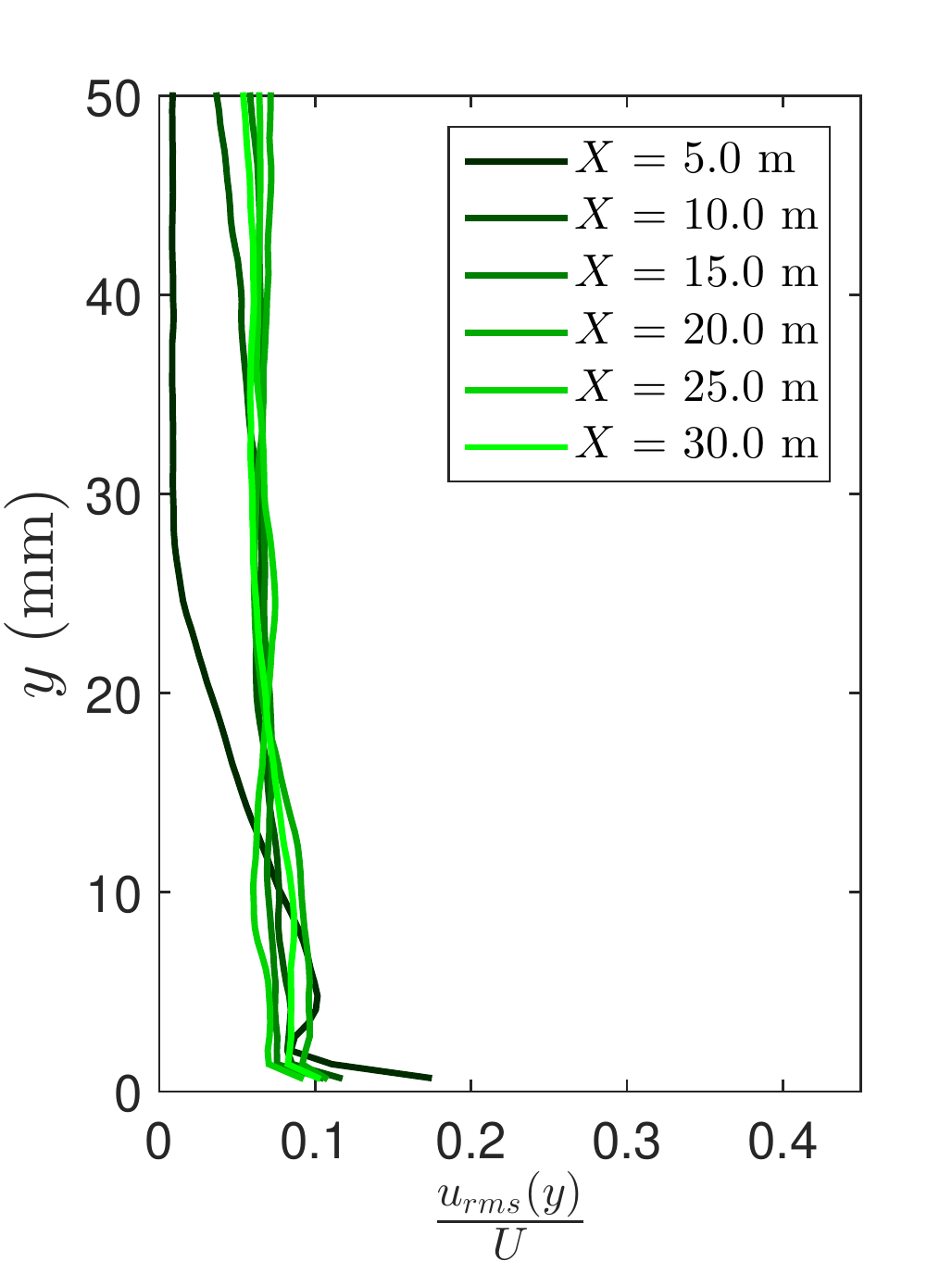} \hspace{.25in} &
\includegraphics[trim=0.5in 0.0in 0in 0.00in,clip=false,width=2.25in]{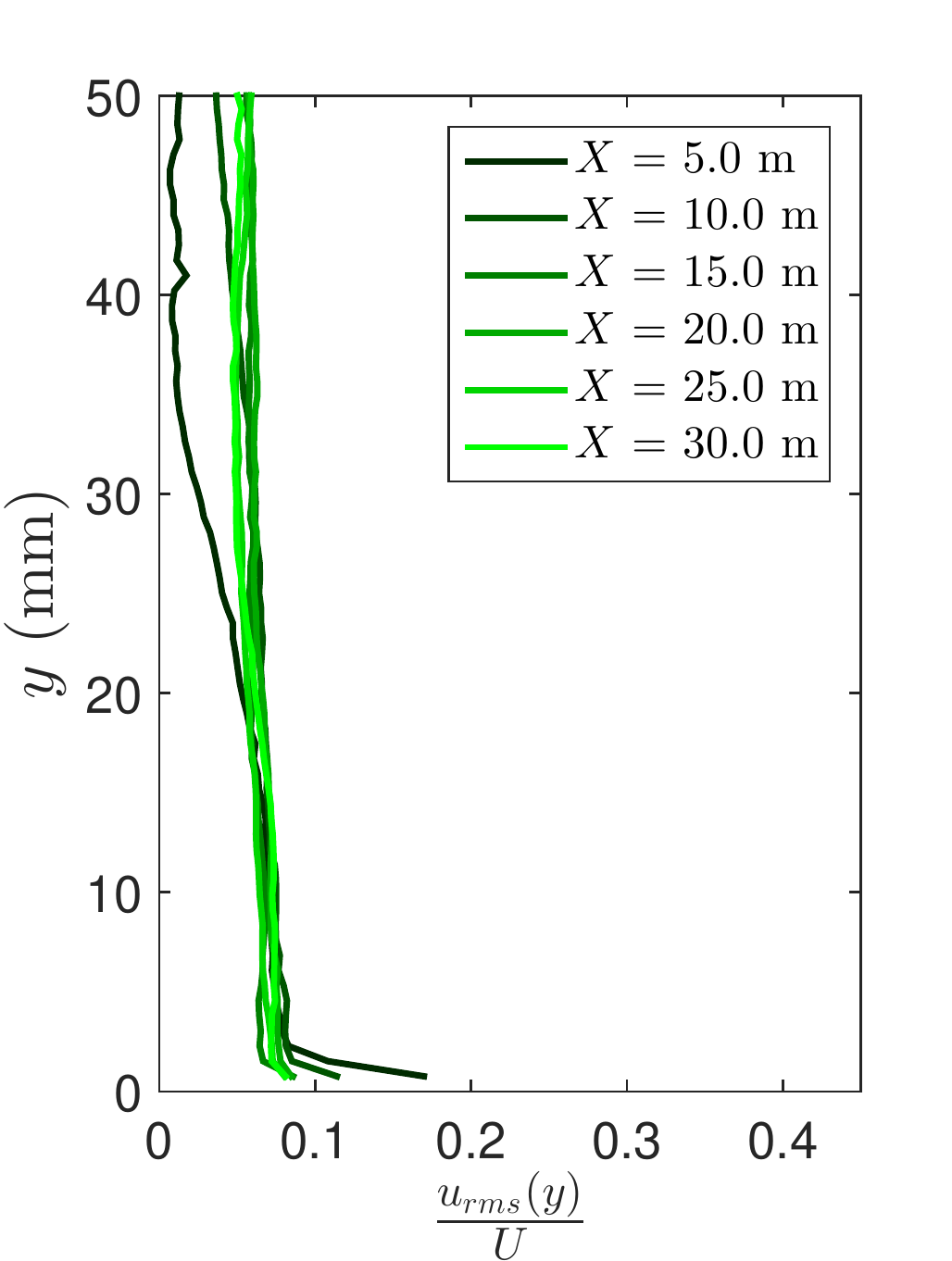}\\
(\textit{c}) & (\textit{d})\\
\end{tabular}
\end{center}
\vspace*{-0.2in} \caption{Streamwise rms velocity profiles at a belt speed of $U$ = 3~m/s for (\textit{a}), (\textit{c}) $D$ = 14~cm and (\textit{b}), (\textit{d}) $D$ = 2.5~cm. These profiles are plotted in (\textit{a}) and (\textit{b}) at each 1~m of belt travel from $x$ = 1 to 5~m and in (\textit{c}) and (\textit{d}) at each 5~m of belt travel from $x$ = 5 to 30~m.} \label{fig:V3_rms_profiles}
\end{figure*}

In addition to mean velocity profiles, additional information can be gained from looking at RMS velocity fluctuations. The RMS velocity at each location is defined as the RMS fluctuation about a mean velocity profile at that $x$ location, discussed in the previous section. At a belt speed of 3~m/s, Figure~\ref{fig:V3_rms_profiles} shows a range of streamwise RMS velocity fluctuations at a range of $x$ values. In this figure, the top row of plots contains profiles from every 1~m of belt travel from $x$ = 1 to 5~m and the bottom row contains profiles at each 5~m of belt travel from $x$ = 5 to 30~m. The left column of plots again shows profiles for the condition $D$ = 14~cm, while the right column of plots shows profiles for $D$ = 2.5~cm. Far from the surface, early profiles have a sharp peak near the belt and decrease away from the belt. After about 2.82~m of belt travel, a second peak in this distribution begins to grow around $y$ = 3 to 4~mm. Close to the surface, a similar sudden burst of activity occurs after only about 1.05~m of belt travel. The appearance of this second peak in the $u_{rms}$ distribution again seems to coincide with the transition to turbulence, seen in the mean velocity profiles as well as the change in growth rate of momentum thickness.

\begin{figure}
\begin{center}
\includegraphics[trim=0.0in 0.0in 0 0.00in,clip=true,width = 3in]{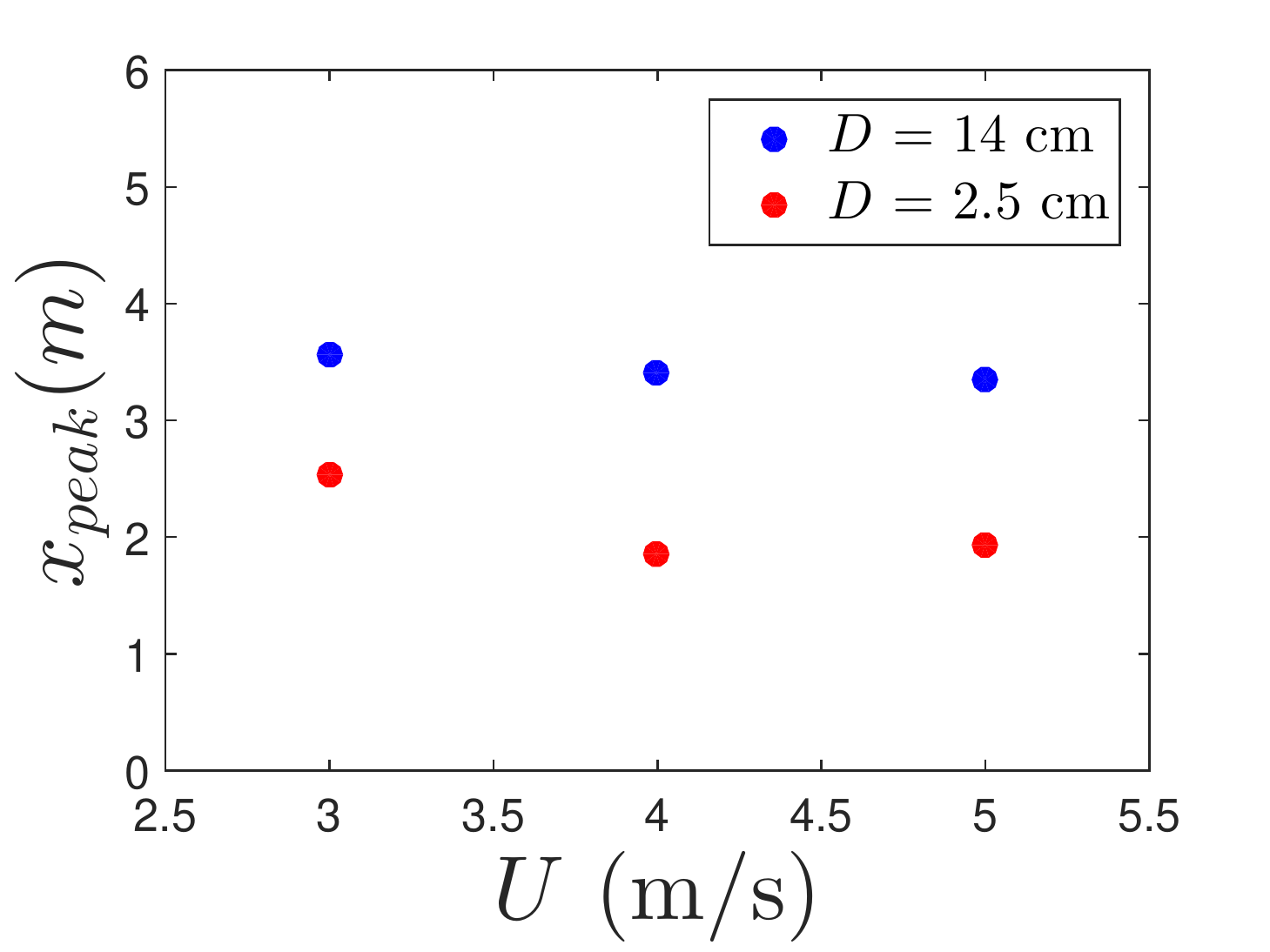} \\
\end{center}
\vspace*{-0.1in} \caption{A plot of the $x$ location at which the secondary peak in streamwise RMS velocity reaches its maximum for each belt speed in each light sheet location.} \label{fig:rms_peak}
\end{figure}

While this second peak begins to form during transition to turbulence, it reaches a maximum value a short time later before flattening out and spreading away from the wall, as can be seen in Figure~\ref{fig:V3_rms_profiles}-(\textit{c}) and (\textit{d}). The $x$ location of this maximum peak can be tracked for each belt speed and light sheet location, as shown in Figure~\ref{fig:rms_peak}.

\subsection{Event Timing}

\begin{figure*}[!p]
\begin{center}
\includegraphics[trim=1in 0.0in 0 0.00in,clip=true,width=6in]{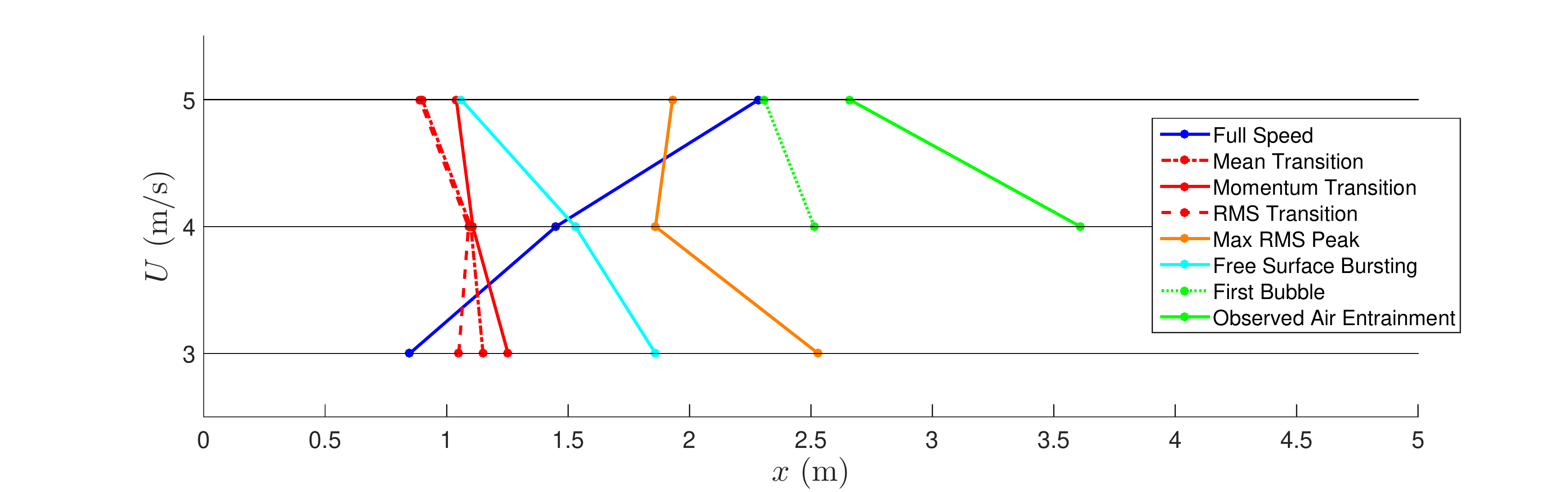}\\
\end{center}
\vspace*{-0.2in} \caption{A set of timelines depicting the $x$ location of various events described throughout this paper.} \label{fig:timeline}
\end{figure*}

The events discussed in the previous sections occur at different times throughout each run. These times correspond to different amounts of belt travel, $x$, depending on the belt speed. In an effort to determine a relationship between these events within each run, while also comparing across different belt speeds, a timeline can be created, as in Figure~\ref{fig:timeline}. In this figure, the horizontal axis converts time to belt travel for each run, normalizing the different timings caused by the acceleration portion of each run. Each horizontal line plots a timeline for each belt speed, with each event type displayed with its own color or line style and connected across belt speeds. Each of the events taken from flow field measurements utilize the near-surface PIV data, as this is thought to be more relevant to free surface bursting and air entrainment. The dark blue line indicates the timing of the belt reaching 95\% of its full speed and appears to have no noticeable correlation to the timing of other events. The three red lines depict the timing of transition to turbulence, each determined using a different indicator, as discussed above. It is clear that the three different indications of transition to turbulence agree fairly well at each belt speed and this transition point appears to move earlier in the timeline with increasing belt speed. The light blue line indicates the timing of free surface bursting. In all cases, this occurs after transition to turbulence, but the distance between transition and bursting seems to shrink with decreasing belt speed, perhaps indicating that the turbulent length or velocity scale is not great enough to disrupt the surface immediately after transition. The two green lines each depict a different method of measuring air entrainment. The dashed green line indicates the location at which air bubbles are first seen to enter the field of view of the camera, using white light movies as discussed in \citet{Washuta2014}. The solid green line indicates the location at which air entrainment is first observed within the field of view of the camera. The second method relies on air entrainment occurring in a very small location along the belt, while the first is not a direct measurement of the onset of air entrainment. While it is unclear which of these methods provides a better indication of air entrainment onset, the timings can both be compared to that of the other events. Little to no air entrainment is observed at a belt speed of 3~m/s. Finally, the orange line indicates the $x$ location at which the secondary peak in streamwise velocity fluctuations reaches a maximum. This peak appears to occur after free surface bursting, but before air entrainment onset. Therefore, while bursting could depend on this velocity scale reaching a critical value, it appears that air entrainment does not depend solely on the magnitude of velocity fluctuations.

\begin{figure*}[!p]
\begin{center}
\begin{tabular}{cc}
\includegraphics[trim=0in 0.0in 0 0.00in,clip=true,width=2.75in]{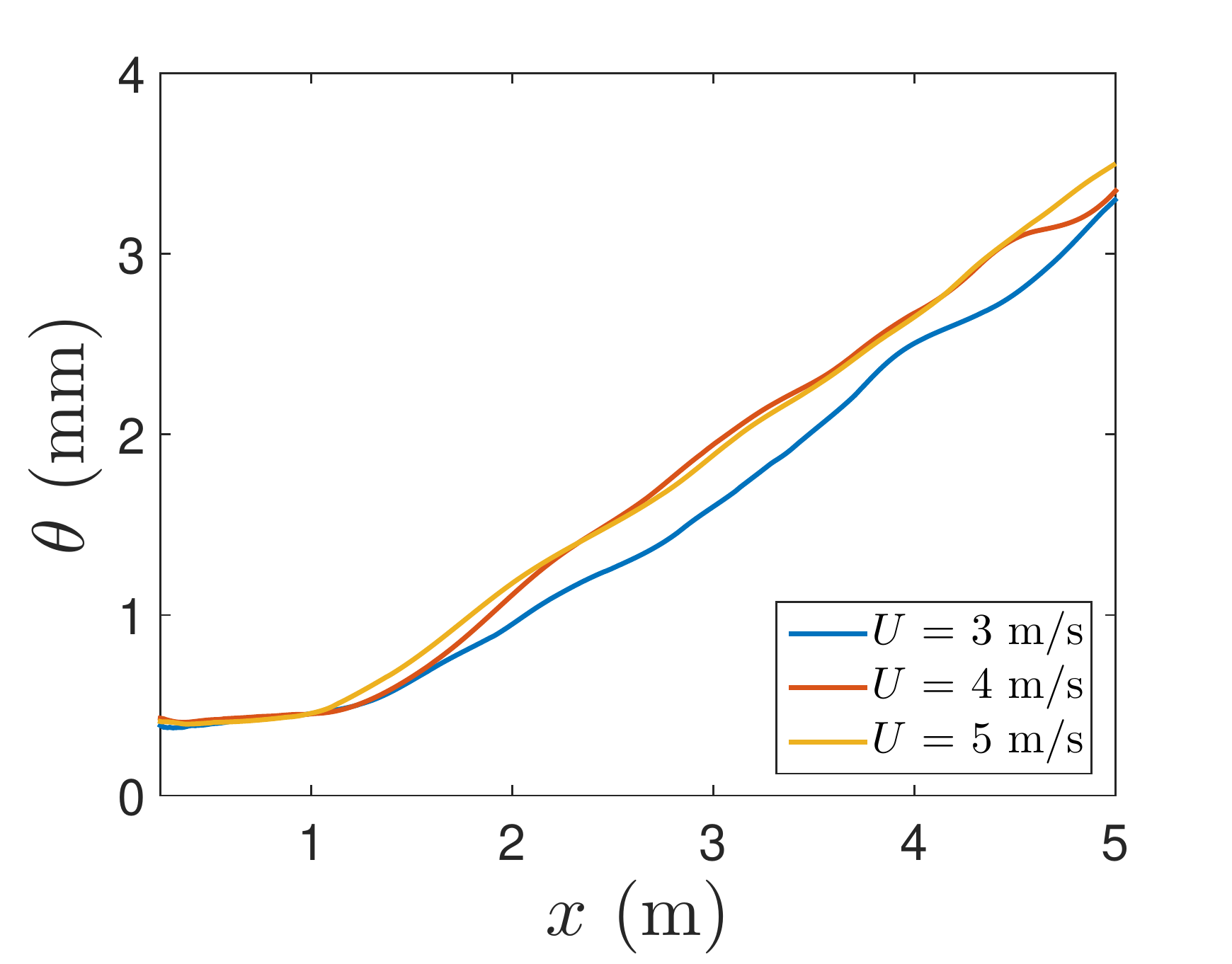}&
\includegraphics[trim=0in 0.0in 0 0.00in,clip=true,width=2.75in]{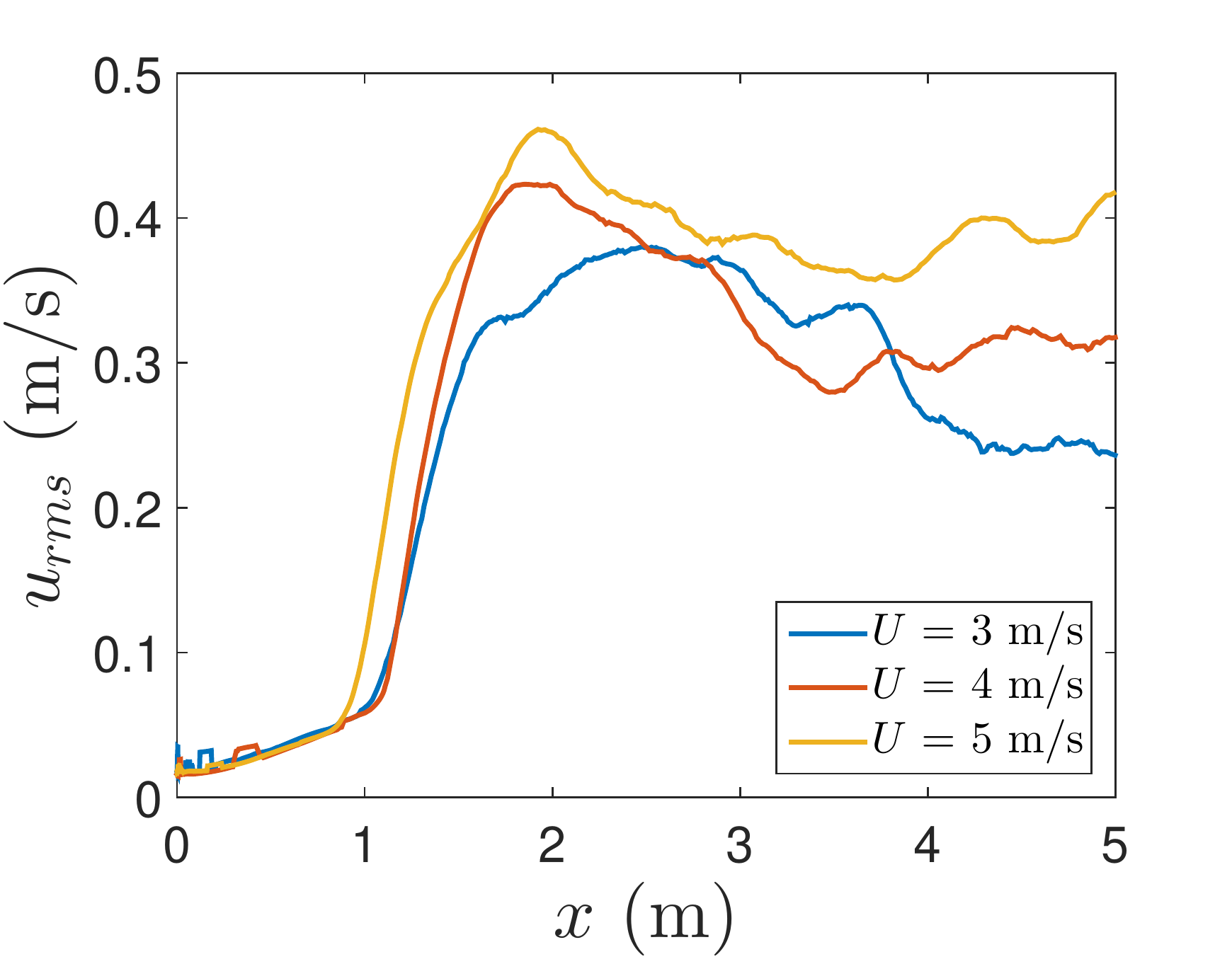}\\
(\textit{a}) & (\textit{b})\\
\end{tabular}
\end{center}
\vspace*{-0.2in} \caption{Plots showing (\textit{a}) $\theta$ vs $x$ and (\textit{b}) $u_{rms}$ vs $x$ for the initial 5~m of belt travel at each belt speed. The data presented in these plots is calculated from velocity measurements averaged over 20 runs.} \label{fig:rms_theta}
\end{figure*}

In order to further study the effect of turbulence on free surface deformations and air entrainment onset, it is important to determine the turbulent length and velocity scales as they evolve throughout each run. As one of the indicators of transition to turbulence, the most obvious length scale that evolves continuously throughout the run is the momentum thickness, $\theta$, shown in Figure~\ref{fig:rms_theta}-(\textit{a}). As a velocity scale, the streamwise velocity fluctuations, $u_{rms}$, in the wall normal location of the secondary peak (see Figures~\ref{fig:V3_rms_profiles}-(\textit{a}) and (\textit{b})) is chosen since it seems to generally coincide with the timings of both free surface bursting and air entrainment. In each run, this secondary peak occurs at $y$ = 3.04~mm and $u_{rms}$ at this location throughout each run is shown in Figure~\ref{fig:rms_theta}-(\textit{b}). These length and velocity scales are calculated from velocity measurements close to the surface. While they evolve differently than those far from the surface or those in typical flat plate boundary layer experiments, these near-surface scales should have more influence on free surface deformations. While the length scale $\theta$ increases steadily throughout the run, this velocity scale reaches a peak and then decreases fairly steadily after that point. This could indicate that air entrainment processes are strongest in this region of rapid boundary layer growth and are reduced later in each run. Additionally, the greater velocity scales reached at higher belt speeds may lend credence to the idea that turbulent velocity fluctuations at lower belt speeds are not great enough to overcome surface tension or gravity in order to entrain air.

\begin{figure*}[!p]
\begin{center}
\begin{tabular}{cc}
\includegraphics[trim=0in 0.0in 0 0.00in,clip=true,width=2.75in]{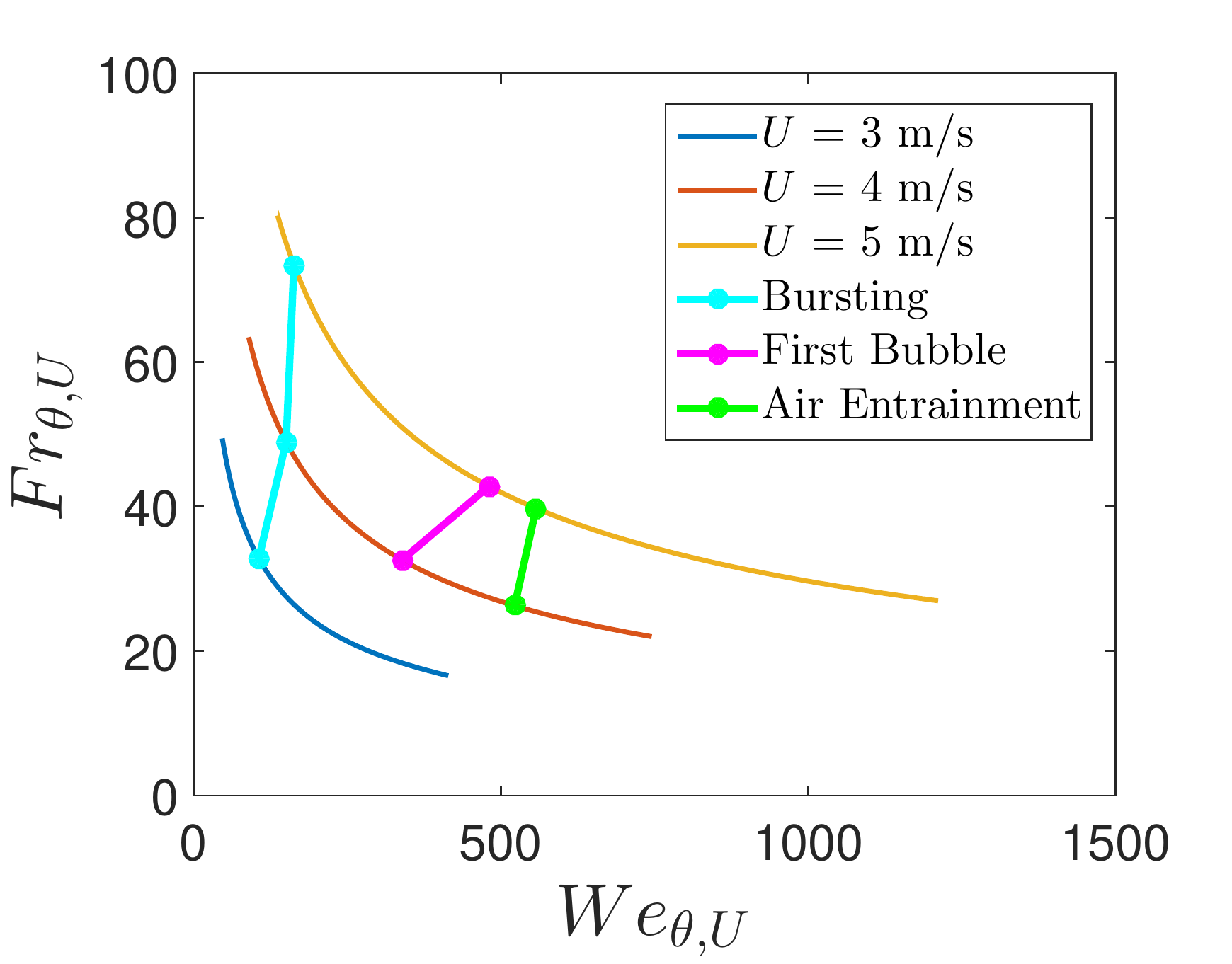}&
\includegraphics[trim=0in 0.0in 0 0.00in,clip=true,width=2.75in]{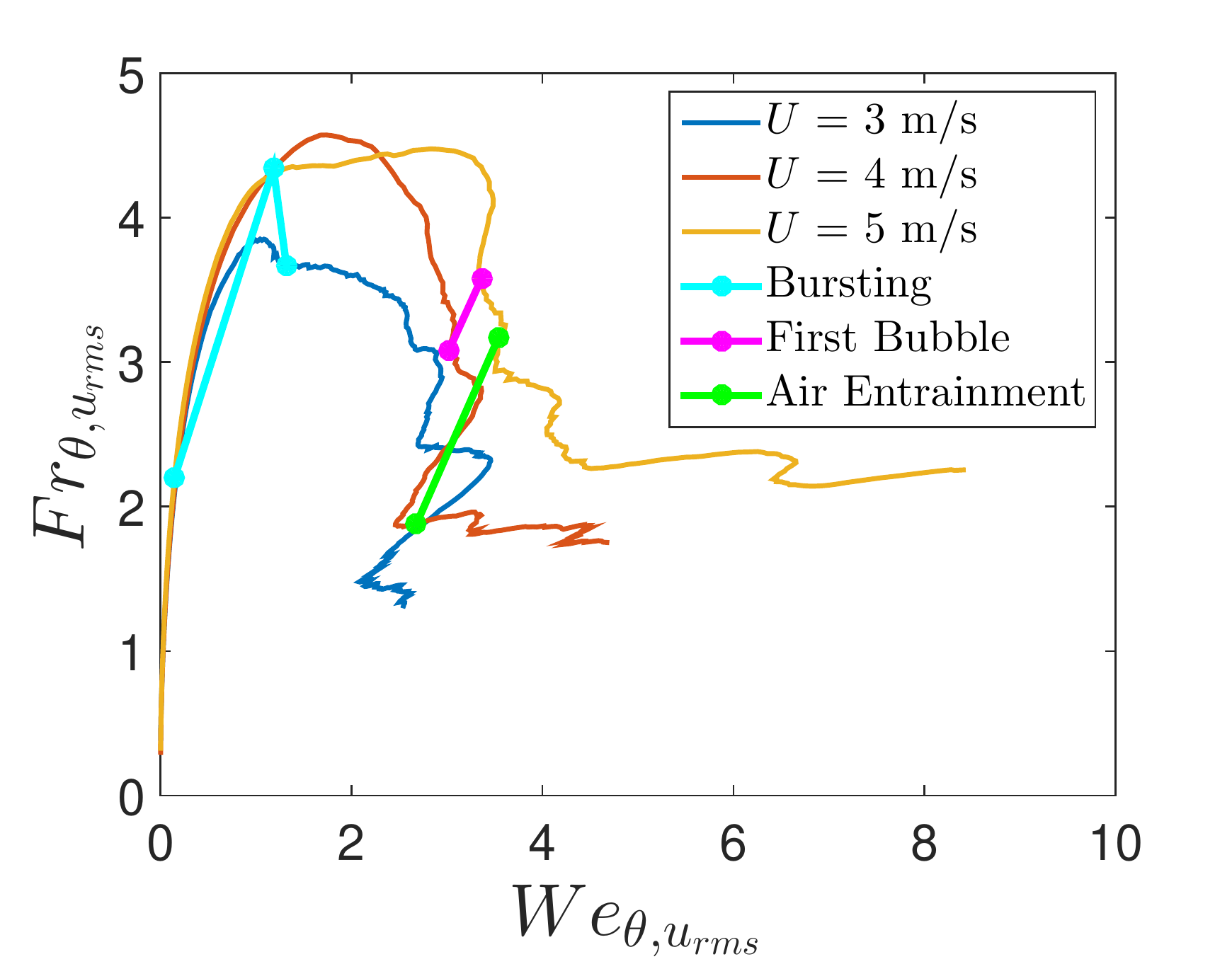}\\
(\textit{a}) & (\textit{b})\\
\end{tabular}
\end{center}
\vspace*{-0.2in} \caption{Plots showing Weber number vs Froude number, calculated using $\theta$ as a length scale and a velocity scale of (\textit{a}) $U$ or (\textit{b}) $u_{rms}$ at $y$ = 3.04~mm, the location of the secondary peak in streamwise velocity fluctuations.  } \label{fig:We_Fr}
\end{figure*}

\begin{figure*}[!htb]
\begin{center}
\includegraphics[trim=2.0in 0.0in 1.5in 0.00in,clip=true,width=5in]{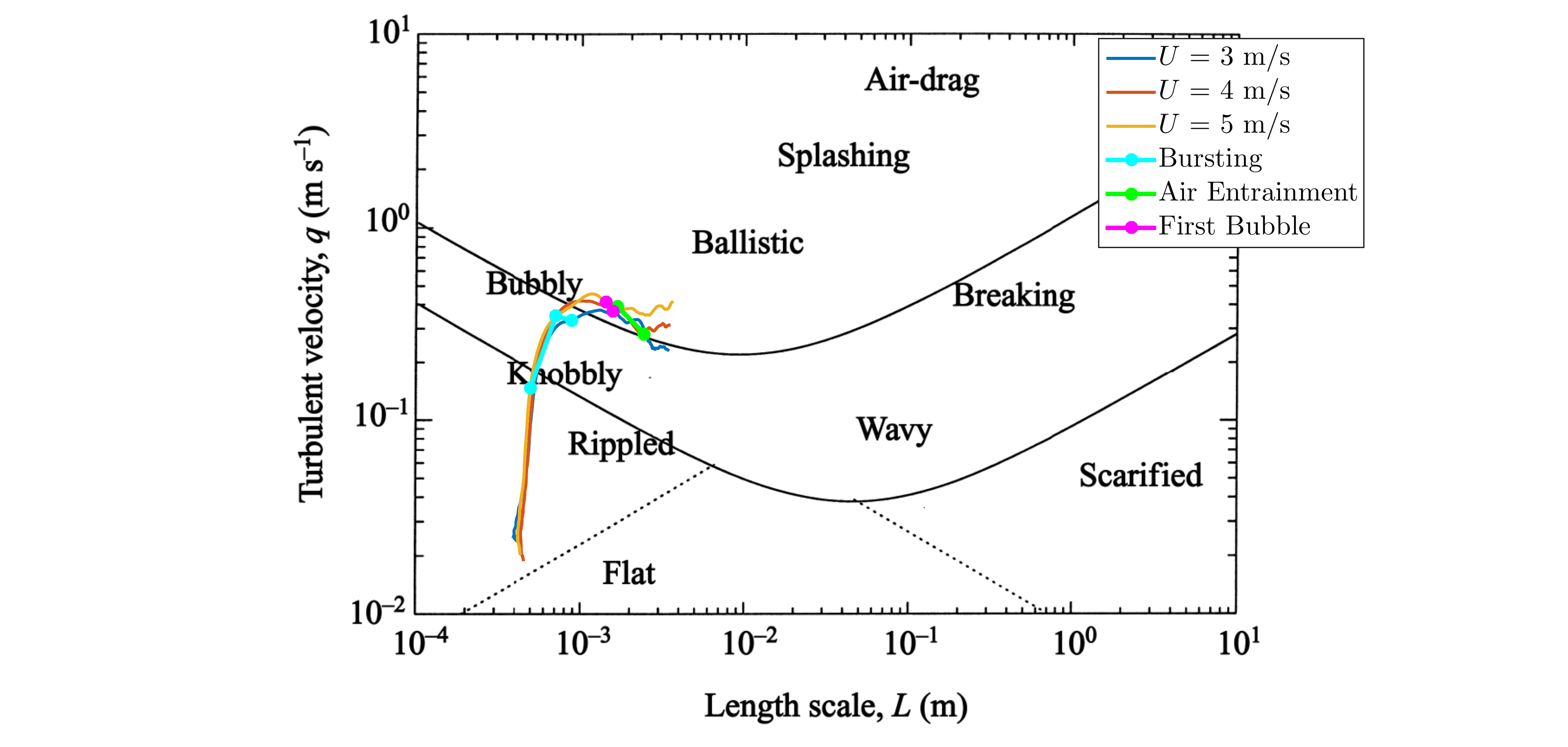}\\
\end{center}
\vspace*{-0.2in} \caption{A figure from \citet{broc:2001} depicting critical regions for air entrainment based on length and velocity scales, with recorded data of $\theta$ and $u_{rms}$ plotted on top. As with the previous figures, belt speeds of 3, 4, and 5~m/s trace paths in blue, red, and yellow, respectively. Light blue indicates the onset location of free surface bursting, while purple and green indicate two methods for determining air entrainment onset.} \label{fig:broc_edit}
\end{figure*}

Using the momentum thickness and the magnitude of the streamwise velocity fluctuations from Figure~\ref{fig:rms_theta}    as velocity and length scales, Weber and Froude numbers can be calculated at each $x$ location and these parameters can be compared across belt speeds for both bursting and air entrainment. Figure~\ref{fig:We_Fr} shows plots of Weber number vs Froude number at each speed, with the values at bursting and air entrainment marked with connected lines, as in Figure~\ref{fig:timeline}. Plot (\textit{a}) shows Weber number and Froude number calculated using $\theta$ as a length scale and the belt speed $U$ as a velocity scale. While Figure~\ref{fig:timeline} depicted increasing $x$ moving from left to right, the plots in  Figure~\ref{fig:We_Fr} trace out a path during the runs of Froude vs Weber number for each belt speed. At small $x$ values, the length scale is small, leading to a small Weber number and large Froude number. As the length of belt travel increases, the momentum thickness grows and the blue, red, and yellow lines trace a path from left to right through this Fr-We space for belt speeds of 3, 4, and 5~m/s, respectively. Using this scaling, free surface bursting appears to depend on reaching a critical Weber number of around 150 in order for onset to occur. This scaling also appears to produce a Weber number threshold of around 500 for air entrainment, shown in green (using the first observed air entrainment method). However, consider that as $x$ increases, the momentum thickness will continue to increase steadily beyond what has been calculated here. Therefore, even at a belt speed of 3~m/s, this threshold will be passed at all three belt speeds. This is in contrast to the observation that no air entrainment occurs at 3~m/s. Therefore, perhaps a more appropriate velocity scale is the $u_{rms}$ of the secondary peak, discussed above. A plot of Weber vs Froude number using $\theta$ as a length scale and $u_{rms}$ as a velocity scale is shown in Figure~\ref{fig:We_Fr}-(\textit{b}) with one curve for each belt speed. While the Weber and Froude numbers in plot (\textit{a}) depended on just $\theta$ and traced a smooth path at each speed, this second plot uses both $\theta$ and $u_{rms}$ to calculate Weber and Froude number. Therefore, at the beginning of each run, both of these values are small and each path begins in the bottom left corner of the plot. As $x$ increases, velocity and length scales both increase and the paths move upward and to the right. As the velocity scale begins to decrease and the length scale continues to grow, the Froude number peaks and begins to decrease, while the Weber number continues to grow. Once again, free surface bursting is depicted in light blue and the two methods of determining air entrainment onset are shown in purple and green. Using this scaling, it is unclear how free surface bursting depends on Weber or Froude number, but air entrainment appears to scale better using these length and velocity scales. From both measurements of air entrainment, it appears that a critical Weber number of approximately 3 to 3.5 must be reached before air entrainment occurs. While this threshold is reached at 3~m/s, it appears to only occur briefly; at higher speeds, the Weber number increases sharply after this point.

An alternative explanation to the idea of a single critical Weber or Froude number for surface bursting or air entrainment is that there is some combined effect of both scaling parameters. This idea is expressed in \citet{broc:2001}, as discussed above and shown in Figure~\ref{fig:brocchiniperegrine}. Using $\theta$ as a length scale and $u_{rms}$ as a velocity scale, as discussed above, these scales can be plotted on top of Figure~\ref{fig:brocchiniperegrine}, as shown in Figure~\ref{fig:broc_edit}. These length and velocity scales are different than the estimates initially provided, but they appear to provide appropriate scales for the turbulence. While the boundary for air entrainment onset does not appear to be appropriate for the length and velocity scales used here, the curves for higher belt speeds do appear to reach further into the upper air entrainment region of the plot, indicating that this figure could be describing the physics correctly.

\section{Conclusions}

In this research, a novel laboratory-scale device was created in order to experimentally study the interaction of the turbulent boundary layer with a free surface. This device utilizes a stainless steel belt, driven by two powered vertically oriented rollers, as a surface piercing vertical wall of infinite length. This belt accelerates in under 0.7 seconds to constant speed $U$ in an effort to mimic the sudden passage of a flat-sided ship. Utilizing the full length and velocity scales of large naval ships, this device creates a temporally-evolving boundary layer analogous to the spatially-evolving boundary layer along the length of a ship, using the transformation $x=Ut$, where $x$ is distance from the leading edge and $t$ is time. Water surface profiles were recorded with a cinematic LIF system to study the generation of surface height fluctuations by the sub-surface turbulence. Sub-surface velocity fields were recorded using a cinematic planar PIV system in order to study the modification of the flow field in the vicinity of the free surface.

It was found that the free surface remains calm during the acceleration portion of the belt launch before bursting with activity close to the belt. This burst location was seen to vary from $x = 1$ to 2~m, depending on belt speed. The Reynolds number based on $x$ at bursting has a consistent value of around $Re_x$ = 5.7 $\times 10^6$. After this point, the free surface fluctuates and changes form rapidly near the belt surface where wave breaking begins to occur. The free surface ripples created near the belt surface appear to lead to the generation of freely-propagating waves far from the belt surface. The speed of these waves normal to the belt surface increases with increasing belt speed, with waves traveling approximately 25\% faster at $U$ = 5~m/s when compared to those at $U$ = 3~m/s.

In studying sub-surface velocity fields, it was found that the boundary layer exhibits similar bursting to what is seen in free surface profiles, with an accompanying secondary peak in streamwise velocity fluctuations, which appears to occur at a $y$ location of approximately 3~mm from the belt surface. The rapid boundary layer growth seen in these velocity profiles happens earlier in planes closer to the surface, leading to an overall thicker boundary layer in the vicinity of the free surface. By analyzing these mean streamwise velocity profiles and comparing the results to the $x$ locations of free surface bursting, it is found that momentum thickness at the location of bursting decreases monotonically with increasing belt speed, while Reynolds number based on momentum thickness provides a fairly consistent indicator of free surface bursting at $Re_{\theta}$ of approximately 1500 based on velocity profiles measured $z=14.0$~cm below the free surface and 2500 based on velocity profiles measured at $z=2.5$~cm.

The onset of both free surface bursting and air entrainment appear to depend in some way on Weber number. Free surface bursting appears to scale well with Weber number of approximately 150 based on momentum thickness and belt velocity, while air entrainment appears to occur after reaching a Weber number of approximately 3 to 3.5 based on momentum thickness and streamwise velocity fluctuations. These scaling parameters appear to agree with the behavior of the scaling arguments presented in \citet{broc:2001}, if not the exact values for this air entrainment boundary.

The support of the Office of Naval Research under grant number N000141110029 (Program Managers: Drs. Patrick Purtell, Ki-Han Kim and Thomas Fu) is gratefully acknowledged.

\bibliography{29ONR}
\bibliographystyle{plainnat}
\setlength{\bibhang}{0pt}

\end{document}